\documentclass[12pt]{article}
\begin{document}
\input epsf
\def\be{\begin{equation}}
\def\bea{\begin{eqnarray}}
\def\ee{\end{equation}}
\def\eea{\end{eqnarray}}
\def\d{\partial}
\def\b{\bigskip}
\def\q{\quad}
\def\d{\partial}
\def\la{\lambda}
\def\eps{\epsilon}
\def\a{{\cal A}}

\begin{flushright}
OHSTPY-HEP-T-01-019\\
hep-th/0109154
\end{flushright}
\vspace{20mm}
\begin{center}
{\LARGE AdS/CFT duality and the black hole information paradox}
\\
\vspace{20mm}
{\bf  Oleg Lunin  and  Samir D. Mathur \\}
\vspace{4mm}
Department of Physics,\\ The Ohio State University,\\ Columbus, OH 43210, USA\\
\vspace{4mm}
\end{center}
\vspace{10mm}
\begin{abstract}

Near-extremal black holes are obtained by exciting the Ramond sector of the
D1-D5 CFT, where the ground state is highly degenerate. We find that the dual
geometries for these ground states have throats that end in a way that is
characterized by  the CFT state. Below the black hole threshold we
find a detailed
agreement between propagation in the throat and excitations of the
CFT. We study
the breakdown of the semiclassical approximation and relate the results to the
proposal of gr-qc/0007011 for resolving the information paradox: semiclassical
evolution breaks
down if hypersurfaces {\it stretch} too much during an evolution. We
find that a
volume ${\cal V}$ stretches to a maximum throat depth of ${\cal V}/2G$.

\end{abstract}
\newpage

\section{Introduction.}
\renewcommand{\theequation}{1.\arabic{equation}}
\setcounter{equation}{0}

One of the most puzzling paradoxes in physics is the black hole
information paradox. When black
holes form and evaporate, the Hawking radiation appears to be
correctly given by a
semiclassical calculation, but the radiation so computed  destroys
unitarity and thus violates the
principles of quantum mechanics.  String theory has made substantial
progress in
understanding the quantum physics of black holes, and its results
suggest very strongly
that the evaporation process maintains unitarity (see for example
\cite{MaldRev,DasMathRev} and references therein). But there is no
clear understanding of the
mechanism by which the information manages to leak out of the black
hole via the radiation.

The essential strength of the information paradox lies in the fact
that Hawking's semiclassical
calculation of the radiation is independent of any details of quantum
gravity at small scales \cite{Hawking}.
Thus even though we find in string theory many perturbative and
nonperturbative quantum
gravity effects at the Planck scale and string scale, it is unclear  how to use
them to effect the information transfer.  It is plausible  that the
effects that correct the
semiclassical computation are nonlocal, and finding them  will
involve fundamental changes in
our understanding of when semiclassical gravity is valid.

            We will work with the D1-D5 system. For this case the microscopic
entropy and
low energy radiation rates agree with the Bekenstein entropy and
Hawking radiation
of the corresponding black hole, including numerical factors
\cite{vs, cm, dm1, dm2}.  Such
agreements contributed to
Maldacena's remarkable conjecture \cite{mal} that the CFT describing the  the
brane system is {\it dual} to
the near horizon geometry produced by the branes. Though this AdS/CFT
correspondence has
been extensively studied, most of the analysis has pertained to the
Neveu-Schwarz (NS) sector
of the CFT, while the black holes in asymptotically flat spacetime
arise in the Ramond (R) sector \cite{henn,exclusion}.
(We will discuss this further below.) The ground state in the NS
sector is unique, but in the R
sector the ground state has a large degeneracy, so that a large
number of microstates appear
to correspond to the same geometry. This fact is not just a technical
complication but an essential
issue, since similar degeneracies yield the Bekenstein entropy of black holes.

In this paper we will analyze the AdS/CFT correspondence in the R
sector. We find several
agreements between microscopic quantities computed in the CFT (which
we take to be a sigma
model at the orbifold point) and the corresponding supergravity
computations. The length
scales  emerging in these relations are quite different from those
that are relevant for the
analysis in the NS sector. The results indicate the way in which the
semiclassical approximation
may break down in string theory to allow information leakage in
Hawking radiation.

\subsection{Review of some basic issues}

Consider first the classical geometry of D1 and D5 branes with no momentum
charge and no angular momentum. At $r\rightarrow\infty$ the
geometry is flat, while at
small $r$ the geometry is locally $AdS_3\times S^3\times M$ where the
4-manifold
$M$ is $T^4$ or K3.   We have a horizon at $r=0$, and the part of
$AdS_3$ covered by the
geometry is one `Poincare patch'. The direction $y$ along the D1-branes
is compactified on a circle of
radius $R$ when we wish to make a black hole in 5 dimensions, and
this creates a periodic
identification of points in the $AdS$ space.

            When we attempt to apply the AdS/CFT correspondence conjecture to
this D1-D5 geometry we
encounter several important questions:

\b

(a) \q The Poincare patch of the AdS geometry smoothly extends past
the horizon $r=0$ to the
entire `global AdS' spacetime. The complete geometry thus obtained
has more than one
asymptotically flat region \cite{spectrum}.  But it appears strange
if new spatial
infinities could be created just
by bringing together a (large but finite) number of branes in a
normal spacetime with
one asymptotic infinity.

One may argue that the semiclassical Poincare patch geometry stops
short of the horizon $r=0$
because the  identification $y\rightarrow y+2\pi R$ relates points that are
closer and closer together as
$r\rightarrow 0$. But this leads to the question: what is the
effective value of $r$ where we
should end the geometry, and what happens when an incident wave
reaches this value of $r$?
One may naively think that the critical value of $r$ would correspond
to the length of the $y$ circle becoming
         order string length or Planck
length. It was argued
however in \cite{lm4} that the geometry ends much further down the
`throat', and that the
truncation of the throat is to be traced to the existence of nonzero
angular momentum in the CFT
state. We will discuss this issue in more detail in this paper.

\b

(b)\q The D1-D5 bound state has a large degeneracy; the number of
ground states is $\sim
e^{2\sqrt{2}\pi\sqrt{n_1n_5}}$, where $n_1$ and  $n_5$ give  the number
of D1 and D5 branes.
The metric of the Poincare patch appears to be the same for all these
states.  This is of course a
version of the standard `black holes have no hair' problem.  One
could try to argue that the branes sit at $r\sim 0$ and thus carry
information about the microstate.
But this would
contradict our
expectation from  the AdS/CFT duality: the duality suggests that the
entire AdS region is
already a dual description of the state of the branes, and we should
not find the branes to be
present at the end of the `throat'.  (We will in fact argue below
that no branes need be included
at the end of the throat; rather the way that the throat ends
characterizes the microstate.)

\b

{{\bf The difference between NS  and R sectors.}}

\b

The initial proposal by Maldacena of the AdS/CFT correspondence
\cite{mal} was motivated
partly by  results on black holes, and pertained to the near horizon
geometry produced by
branes in asymptotically flat spacetime. When we wrap D1,  D5 branes
on a circle $y$ in
spacetime, then the behavior of the fermions in the CFT is induced
from the behavior of
fermionic fields in the bulk. Since the supergravity in the bulk has
fermions periodic around
$y$, the D1-D5 CFT also has fermions that are periodic around the
$y$ circle, so the CFT is in
the R sector.  (If we put the fermions in the bulk supergravity to be
antiperiodic around $y$
then the vacuum energy is nonzero, and flat spacetime ceases to be a
solution. Since we want
asymptotically flat spacetime, we will not consider this situation.)

The proposal in \cite{mal} did not however give an explicit way to
relate correlators via the
duality. The map of chiral operators and correlators was carried out
by Gubser, Klebanov,
Polyakov \cite{gkp} and by Witten
\cite{witten}.  Let us analyze these and other approaches to AdS
correlators, keeping in mind
that we are interested in the following essential physics. A wave
traveling towards $r=0$
becomes one of shorter and shorter wavelength, due to the redshift
between infinity and the
small $r$ region. The wave does not reflect back to larger $r$ unless
we have some explicit
modification to the physics of the `throat'. This monotonic infall of
an incident quantum to the
horizon is an essential aspect of the black hole problem; the
particle itself does not appear to
return but the deformation it creates in the geometry causes Hawking
radiation to be emitted,
and this radiation appears to carry no information about the infalling quantum.

In the analysis of Witten \cite{witten} the AdS space was rotated to
Euclidean signature. This
causes the Poincare patch to become a smooth space with no
singularity at $r=0$. However at
the same time we lose the physics of the horizon; we cannot have
traveling waves in Euclidean
signature, nor do we have the accumulation of wavefronts near $r=0$
that signals the infall of
the particle to the horizon. The situation is similar to rotating a
black hole geometry to Euclidean
signature: we get only the space outside the horizon, and the
geometry is smooth at the location
$r=r_{horizon}$. But with such a rotated geometry we cannot address
the question of what
happens to quanta that fall into the hole.

In \cite{gkp} correlation functions were computed in the Poincare
patch arising from the near
horizon geometry of branes. The correlators were however
computed for {\it spacelike} momenta $p^2>0$. The wavefunction in
this case has a growing part
and a decaying part at $r=0$, and the growing part was set to zero to
solve for the Green's
function. But real infalling particles have timelike or null momenta.
The waveform
is then oscillatory near $r=0$, and we have no natural way to relate
the ingoing and outgoing
parts of the wave. (If we just drop the outgoing part, then we have
particles being swallowed
into the horizon, and we cannot make a Green's function at the AdS
boundary by using the
propagator.)

Correlation functions have also been computed in the AdS geometry with
Lorentzian signature and with arbitrary momenta, but in these
calculations the spacetime was
extended to the `global AdS' space past
$r=0$ \cite{balas}.  To relate computations with this global AdS to the
black hole
question that we started with, we would have to connect the global
AdS to asymptotically flat
spacetime. It is unclear how this is to be done, especially if we
wish to avoid the appearance of
new asymptotically flat regions. As was mentioned above, it would be
strange if matter falling
onto the branes were to move smoothly through $r=0$ and emerge in a
new asymptotically flat
region, since we could make the brane geometry by gathering together
branes in a spacetime
which started with only one asymptotic infinity.

Thus we see that we cannot directly use any of the usual ways of
computing correlators in the
AdS/CFT correspondence to understand how information may return (as
Hawking radiation)
after we throw some energy into the throat of the D1-D5 brane
geometry. As mentioned before
the difficulty is not just a technical issue, but rather the fact
that if we could set up AdS
correlators in a supergravity geometry that pertains to a black hole
then we would directly
observe information emerging in Hawking radiation. But extensive work
with supergravity
solutions have shown that we cannot get such a simple resolution of
the black hole information
problem; further the emerging Hawking radiation is a complicated
multi-particle state even for a
single high energy incident particle, so it makes sense that we are
unable to compute simple
2-point functions in AdS  that correspond to returning a particle from $r=0$.

To address the black hole information problem we need to work in the
R sector of the CFT, with
Lorentzian signature, an asymptotically flat infinity, and with
quanta that have timelike or null momenta. The global AdS
geometry corresponds to the NS sector of the CFT; we must instead use
the Poincare patch and
try to resolve the physics at $r=0$.
Changing any of these conditions  bypasses the information  problem, even
though we may get interesting AdS/CFT correspondence theorems.

\subsection{Results}

(A) \q We start by listing the R ground states of the D1-D5 CFT. These states
may be represented pictorially in terms of the `effective string',
which has total
winding number $n_1n_5\equiv N$      around the $y$ circle. In a generic
state    the effective string is broken up into a number $m$ of
       `component strings'; each component string is wrapped $n_i$ times
around $y$ before closing
       on itself, and $\sum _i n_i = N$.

       Naively, the geometry of the D1-D5 supergravity solution appears to
have a uniform throat
       (in the sense that that angular $S^3$ asymptotes to a constant
radius), all the
       way down to a horizon at $r=0$. We find however that the throat ends
at some point
       before $r=0$. There is a singularity at the end of the throat
characterized by a curve in
       4-d space. The location and      shape  of this curve mirrors the
       CFT microstate.   In the special case that the curve is a circle we
recover the metrics of
       \cite{mm}.

             \b

       (B)\q If the throat extended all the way down to $r=0$ then
throwing  even an
infinitesimal energy into the throat would lead to horizon formation
at some value of $r$.
Since the throat is actually finite, there is an energy scale $\Delta
E_{threshold}$ below which
we can study matter quanta moving in the throat without formation of
a horizon (this is the
`hot tube' studied in \cite{lm4}).

Below the threshold of black hole formation  $\Delta
E_{threshold}$ we construct a detailed map between the throat
geometry and the dual CFT
effective string. A quantum placed in the throat bounces several
times up and down the
throat before escaping to infinity; let the time for each bounce be
$\Delta t_{SUGRA}$. In the
dual CFT this quantum is represented by a set of left and a set of
right movers on the effective
string; these vibration modes travel around the string and meet each
other at intervals
$\Delta t_{CFT}$.  For the special metrics of \cite{mm} we can
separate the wave equation and
obtain
$\Delta t_{SUGRA}$ precisely \cite{lm4}.  We find

\b

(i)\q $\Delta t_{CFT}=\Delta t_{SUGRA}$

\b

(ii)\q The probability per unit time for the quantum to escape from
the supergravity throat
exactly equals the probability per unit time for the vibration modes
on the effective string to
collide and emerge as a graviton.

\b

We then proceed to argue that even though we cannot solve the wave
equation for a general
microstate, we still get $\Delta t_{CFT}\sim \Delta t_{SUGRA}$ for
all microstates.

Note that the  scale $\Delta t_{SUGRA}$ is much larger than the radius of the
$AdS_3$ which appears in $AdS$/CFT computations of correlation
functions performed in the
NS sector \cite{gkp,witten, freedman}; thus we are probing a different set
of quantities than are
usually studied with the
duality. Also note that in comparing  emission rates we are using
explicitly the part of
spacetime that joins the near horizon geometry to flat space.

\b

(C)\q  In setting up the map between quanta in the supergravity throat and
       vibration modes of the CFT component strings we find the following:
       the naive map breaks down if we are forced to put more than one set of
vibration modes
       on the same component string.   In general the CFT state has several
component strings,
        and
       we can thus describe several quanta in the supergravity throat.
Interestingly, when we put
       enough quanta that we would run out of component strings in the CFT,
we find on the
       supergravity side that we reach the threshold $\Delta E_{threshold}$ for
black hole
       formation.

       At this point we note a close similarity of these results with a
conjecture
       made in \cite{mes}.    In \cite{mes} it was argued that to resolve the
black hole information paradox
       we need spacetime to have the following property: semiclassical
propagation
on the spacetime
       breaks down if during the course of evolution an initial slice in a
foliation is {\it stretched}
       beyond a certain length. Thus spacelike slices need to be 
endowed not only
with their
       intrinsic geometry but also with a `density of degrees of 
freedom'; if the
stretching
       dilutes these degrees too much then nonlocal effects spoil the usual
evolution equation.

        We find that the number of component strings in the CFT state
act like the
        density of degrees of freedom for the dual throat geometry -- when
        space is stretched to give longer throats we also have fewer component
strings.
We characterize the  `stretch'    of a volume ${\cal V}$ by the time
$\Delta t$ to travel down and back up the throat.
We then find the relation
\be
\Delta t_{max} ={{\cal V}\over 2 G_N^{(5)}}
\ee

\section{R ground states in the CFT and  dual metrics for a
special family}
\renewcommand{\theequation}{2.\arabic{equation}}
\setcounter{equation}{0}

\subsection{The states in the CFT}

We consider the bound states of $n_1$ D1 branes and $n_5$ D5 branes
in IIB string theory. We
set
\be
N=n_1n_5
\ee
The D5 branes are wrapped on a 4-manifold $M$, and thus appear as
effective strings in the
remaining 6 spacetime dimensions. $M$ can be $T^4$ or K3. The  D1 branes
and the effective strings from the D5 branes extend along a common
spatial direction $x_5\equiv y$, and
            $y$ is compactified on a circle of length $2\pi R$.  The low
energy dynamics of this system is a
{\cal N}=(4,4) supersymmetric 1+1 dimensional conformal field theory (CFT).
The CFT has an internal R-symmetry $SU(2)_L\times SU(2)_R\approx SO(4)$.
This
symmetry arises from the rotational symmetry of the brane
configuration in the noncompact
spatial directions $x_1$, $x_2$, $x_3$, $x_4$. The group $SU(2)_L$ is
carried by the left movers in the
CFT and the group $SU(2)_R$ is carried by the right movers.

Consider this CFT at the `orbifold point' \cite{sw, db, dj, lm,
wadia}. Then the CFT is a 1+1
dimensional sigma model where
the target space is the orbifold $M^N/S_N$, the symmetric product of
$N$ copies of the
4-manifold $M$. We are interested in the R sector ground states of
this system (these will be
states with no left or right moving excitations).

These R ground states can be obtained by first finding all the chiral
primary fields in the NS
sector, which are states with $h=j_3$, $\bar
h=\bar j_3$.  Spectral flow then maps these chiral primaries to ground
states of the R sector.
          Spectral flow acts
independently on the left and right movers, in the following way
\bea
h^R&=&h^{NS}-j_3^{NS}+{c\over 24}\nonumber \\
j^R_3&=&j^{NS}_3-{c\over 12}
\label{two}
\eea
The R ground states all have $h=\bar h={c\over 24}$.

\b

{\subsubsection{Chiral primaries in the NS sector.}}

\b

The $M^N/S^N$ orbifold CFT and its states can be understood in the
following way.  We take $N$
copies of the supersymmetric $c=6$ CFT which arises from the sigma
model with target space
$M$. The vacuum of the theory is just the product of the vacuum in
each copy of the CFT.  In
the  orbifold theory we find twist operators  $\sigma_n$ \cite{dvv,
jev}. The copies
$1, 2,
\dots n$ of the CFT  permute  cyclically into each other
$1\rightarrow 2\rightarrow
\dots
\rightarrow n
\rightarrow 1$ as we circle the point of insertion of $\sigma_n$.
(The other copies are not
touched, and we ignore them for the moment.) In this given twist
sector there are operators
with various values of
$j_3$, but we are interested in those  that are chiral operators. The
chiral operator in this twist
sector with lowest dimension and charge is termed
$\sigma_n^{--}$ \cite{lm2} and has
\be
h= j_3={n-1\over 2}, ~~\bar h= \bar j_3={n-1\over 2}
\ee
Each copy of the CFT has the SU(2) currents $J^{(i)a}, \bar J^{(i)a}$,
where the index $i$ labels the
copies.  Define
\be
J^a=\sum_{i=1}^n J^{(i) a}, ~~~ \bar J^a=\sum_{i=1}^n \bar J^{(i) a}
\ee
Then we can make three additional chiral primaries from $\sigma^{--}$:
\bea
\sigma_n^{+-}&=&J_{-1}^+\sigma_n^{--}, ~~~~~~~h=j_3={n+1\over 2}, ~\bar
h=\bar j_3={n-1\over
2}\nonumber \\
\sigma_n^{-+}&=&\bar J_{-1}^+\sigma_n^{--}, ~~~~~~~h=j_3={n-1\over 2},
~\bar h=\bar j_3={n+1\over
2}\nonumber \\
\sigma_n^{++}&=&J_{-1}^+\bar J_{-1}^+\sigma_n^{--}, ~~~h=j_3={n+1\over 2},
~\bar h=\bar j_3={n+1\over
2}
\eea
The chiral primaries $\sigma_n^{--}, \sigma_n^{+-}, \sigma_n^{-+},
\sigma_n^{++}$ correspond respectively
to  the
$(0,0)$, $(2,0)$,  $(0,2)$, $(2,2)$ forms from the cohomology of $M$. Both
$T^4$ and K3 have one form
of each of these degrees. We will not consider the chiral primaries
arising from
the other forms on $M$; this will not affect the nature of the
arguments that we wish to present.

The operator $\sigma_1^{--}$ is just the identity operator in one
copy of the $c=6$ CFT. Thus for
the complete CFT made from $N$ copies we can write the above chiral
operators as
\be
\sigma_n^{\pm\pm}[\sigma_1^{--}]^{N-n}
\ee
It is understood here that we must symmetrize the above expression
among all permutations
of the $N$ copies of the CFT; we will not explicitly mention this
symmetrization in what follows.

More generally we can make the chiral operators
\be
\prod_{i=1}^k ~[~\sigma_{n_i}^{s_i, \bar s_i}~]^{m_i},
~~~~~~\sum_{i=1}^k n_im_i=N
\label{one}
\ee
where $s_i, \bar s_i$ can be $+,-$. This
gives the complete set of chiral primaries that result if we restrict
ourselves to the above
mentioned cohomology of
$M$.
\b

{\subsubsection{Ground states in the R sector.}}

In the state (\ref{one}) the $N$ copies of the $c=6$ CFT are
naturally grouped into subsets of size
$n_i$.  The operation of spectral flow proceeds
independently in each such
subset.  Thus consider a subset corresponding to the $n$ copies
linked by a twist $\sigma_n^{s,
\bar s}$. For this subset we have $c=6n$ in the spectral flow
relations (\ref{two}).  After the flow
we get the charges
\bea
\sigma_n^{--}&\rightarrow &j_3^R=-{1\over 2}, ~\bar j_3^R=-{1\over
2}\nonumber \\
\sigma_n^{+-}&\rightarrow &j_3^R={1\over 2}, ~~~\bar j_3^R=-{1\over
2}\nonumber \\
\sigma_n^{-+}&\rightarrow &j_3^R=-{1\over 2}, ~\bar j_3^R={1\over
2}\nonumber \\
\sigma_n^{++}&\rightarrow &j_3^R={1\over 2}, ~~~\bar j_3^R={1\over 2}
\label{three}
\eea
In fact these 4 operators, after flowing to the R sector, join up to
form  a representation
of $SU(2)_L\times SU(2)_R$ with $(j, \bar j)=({1\over 2}, {1\over
2})$. This can be seen by
noting that $J_{-1}^+$ in the NS sector flows to $J_0^+$ in the R sector.

This treatment of each separate $\sigma_n^{s, \bar s}$ would complete
our treatment of the R
ground state except for the fact that when we have more than one copy
of the same operator
then we must take only the symmetric combination of the different
copies. Thus consider the
set of chiral primaries (\ref{one}) where the twist sector $n$ occurs
a total of $m$ times, and
concentrate attention on the $nm$ copies of the $c=6$ CFT that are
involved in  these
operators.

In the absence of symmetrization we would have, after flowing to the
R sector, spin values $(j,
\bar j)=({1\over 2}, {1\over 2})$ from each set (\ref{three}), and we
would add the spins
according to the rules for SU(2) independently in the left and right
SU(2)s. The symmetrization
has the effect of giving, in the R sector, only the  states with
equal values of $j_L$ and $j_R$:
\bea
(j,j^3;j,{\bar  j}^3), ~~~j&=&{m\over 2}, {m\over 2}-1, \dots ,0;~~
~~m~{\rm  even},\nonumber
\\ j &=&{m\over 2}, {m\over 2}-1, \dots ,{1\over 2}; ~~ ~~m~{\rm  odd}.
\label{nine}
\eea
(The values of $j^3, {\bar j}^3 $
range independently from $-j $ to  $ j$.)

\b

{\subsubsection{Pictorial description of R ground states.}}

\begin{figure}
\begin{tabular}{ccc}
\epsfysize=1.5in \epsffile{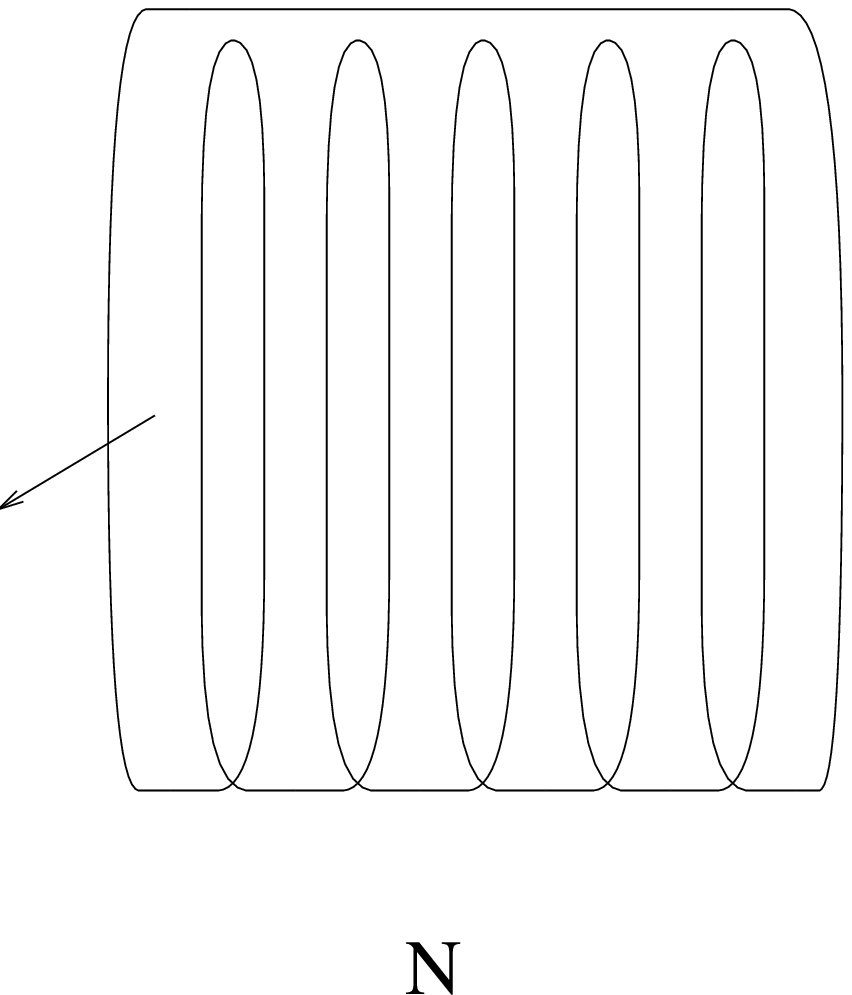}&
\begin{picture}(80.00,10.00)
\end{picture}&
\epsfysize=1.5in \epsffile{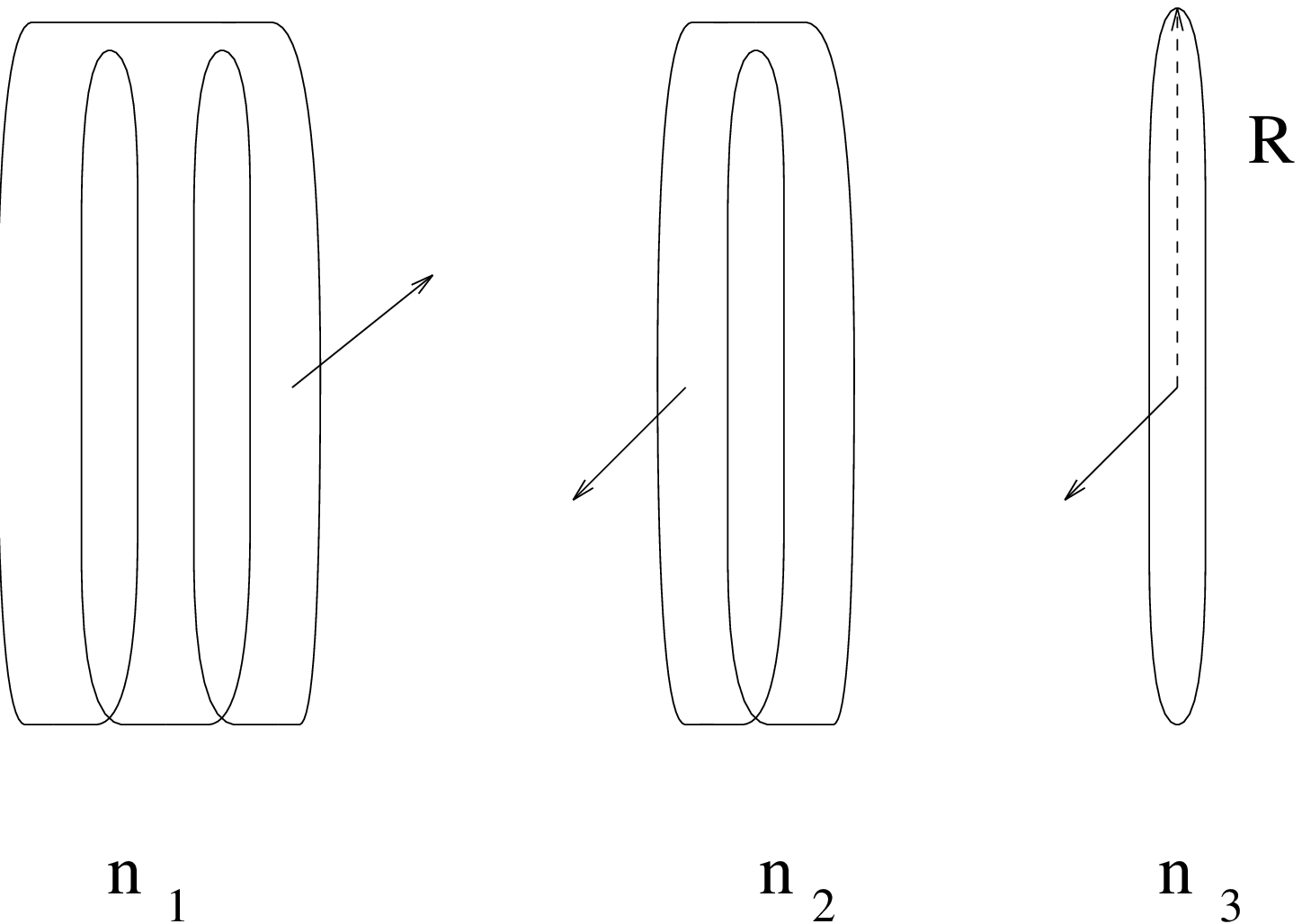}\\
(a)&&(b)
\end{tabular}
\caption{\label{figGround} (a) The state with one component string
wrapped $n_1n_5$ times around
the direction $y$; \q (b)
A generic state with several component strings. The arrows
indicate the spins of the component strings.}
\end{figure}

The twist operator $\sigma_N^{s, \bar s}$ joins all $N$ copies of the
$c=6$ CFT into one copy of the
$c=6$ CFT (on an $N$ times longer spatial circle).  We  depict the
corresponding  R ground state
pictorially in figure \ref{figGround} (a); we  have a `multiwound' string
wrapped
$N=n_1n_5$ times around the
$y$ circle. A generic state of the form (\ref{one}) is pictured in
figure \ref{figGround}(b); there are $k$
`component strings', with the $i$th string  wrapped $n_i$ times
around the $y$ circle.

Each component string carries spin $(j_L, j_R)=({1\over 2}, {1\over
2})$, which we have
depicted by the arrows on the component strings.  We get a
particularly simple set of
geometries if all the spins are aligned (these geometries will be
reviewed in the subsection
below), but we will need to consider  general spin orientations to
address the physics of the
generic microstate.

\subsection{A special family of metrics for the D1-D5 system}

Let us recall the physics of a special family of metrics for the
D1-D5 system; these metrics were
studied in \cite{mm,lm4}.  We will see in the next section that these
special metrics correspond to CFT
states arising from chiral primaries $[\sigma_{N/m}^{--}]^m$; i.e.,
we have many component
strings of equal length, and their spins are all aligned. The D1
and D5 charges are $Q_1, Q_5$ respectively, and
the angular momentum is specified by a parameter $a$. The 10-D metric
and other supergravity
fields are given in Appendix \ref{AppD1D5Rev}, but for the most part
we will need
only the 6-D Einstein
metric obtained by dimensional
reduction on the 4-manifold $M$:
\bea\label{MaldToCompare}
d{s}_E^2&=&-\frac{1}{h}(d{t}^2-d{ y}^2)+
hf\left(d\theta^2+\frac{d{r}^2}{{r}^2+a^2}\right)
-\frac{2a\sqrt{Q_1 Q_5}}{hf}\left(\cos^2\theta d{ y}d\psi+
\sin^2\theta d{ t}d\phi\right)\nonumber\\
&+&h\left[
\left({ r}^2+\frac{a^2Q_1Q_5\cos^2\theta}{h^2f^2}\right)
\cos^2\theta d\psi^2+
\left({ r}^2+a^2-\frac{a^2Q_1Q_5\sin^2\theta}{h^2f^2}\right)
\sin^2\theta d\phi^2\right],\nonumber \\
\eea
where
\be\label{defFHProp}
f={r}^2+a^2\cos^2\theta,\qquad
h=\left[\left(1+\frac{Q_1}{f}\right)\left(1+\frac{Q_5}{f}\right)\right]^{1/2}
\ee

This geometry is flat at spatial infinity.  The direction $y$ is
assumed to be compactified on a
circle of radius $R$. The size of the
$S^3$ in the variables
$\theta,
\phi,
\psi$ settles down to a constant for $a\ll r\ll (Q_1Q_5)^{1/4}$, so we
term the region
$r<(Q_1Q_5)^{1/4}$ as a `throat'.  Note however that since the $y$
circle shrinks as $r$
decreases, we will not in fact get a uniform throat if we
dimensionally reduce along $y$ and
look at the 5-D Einstein metric. Thus we do not have a throat in the
same sense as we get for
the D1-D5-momentum system, but we continue to use the term `throat' since it
conveys the constancy of the $S^3$ radius. The start of the throat is
at $r\sim (Q_1Q_5)^{1/4}$,
but it is important that the throat is `capped off' at  $r\sim a$:
the geometry ends
smoothly except for  a singularity on the curve $r=0,
\theta=\pi/2$.

Consider the massless scalar wave equation  in the above metric. We
can   write the anzatz
\be
\Phi(t,r,\theta,\phi,\psi,y)=\exp(-i\omega t+im\phi+in\psi+i\la y)
{\tilde\Phi}(r,\theta).
\ee
It turns out however that there is an additional hidden symmetry in
the metric, and there is a
further  separation between $r$ and
$\theta$ \cite{larsen}.

In \cite{lm4} the following process was studied. We start with  a
quantum of this scalar field at
spatial infinity, sent in towards $r=0$.  Let the wavelength
$\omega^{-1}$ be much larger than
the length scale
$(Q_1Q_5)^{1/4}$. Then most of the waveform reflects back to infinity
from the start of the
throat, but there is a small probability $P$ for the quantum to enter
the throat \cite{ms1, lm4}
\be
P=4\pi^2\left(\frac{Q_1Q_5\omega^4 }{16}\right)^{l+1}
\left[\frac{1}{(l+1)!l!}\right]^2
\label{sone}
\ee
where $l$ specifies the spherical harmonic of the wave.

            The part of the wave which enters the throat travels to
$r=0$, where it
reflects back;  we find naturally reflecting boundary conditions at
the singularity.  When the wave reaches
back to the start of the
throat we again have the same probability $P$ that the quantum will
escape to infinity, while the
probability is $1-P\approx 1$ that it will reflect back into the
throat.  Thus the quantum travels
several times in the throat before escaping. We can find the time for
traveling once up and
down the throat by looking at the phase shift between successive
wavepackets emerging from
the throat. This time is
\be
\Delta t=\pi\frac{\sqrt{Q_1Q_5}}{a}.
\label{tone}
\ee

The parameter $a$ is related to the angular momentum of the geometry
as follows. The rotation
group on the noncompact spatial directions is $SO(4)\sim SU(2)\times
SU(2)$.  The value of $j$
in each $SU(2)$ factor is the same,  and it can be an integer or half
integer. We have
\be
a={2j\over n_1n_5} {\sqrt{Q_1Q_5}\over R}
\label{ttwo}
\ee
Following \cite{mm} we set
\be
\gamma\equiv {2j\over n_1n_5}
\label{stwo}
\ee
The maximum value of $j$ is ${n_1n_5\over 2}$, so the maximum value
of $\gamma$ is unity.

Substituting (\ref{ttwo}) in (\ref{tone}) we get
\be
\Delta t={\pi R\over \gamma}
\label{tthree}
\ee

\section{ CFT states and their dual   supergravity throats}
\renewcommand{\theequation}{3.\arabic{equation}}
\setcounter{equation}{0}
\label{secthree}

\b

{\subsection{Microscopic model used for  black hole absorption computations.}}

We have seen in the past that absorption cross sections for  D1-D5
black holes  can be
reproduced, including numerical factors, by the computation of
absorption into the
corresponding microstate. Let us briefly recall the nature of these
microscopic computations.

The microststate   was described by an `effective string' wound
$N=n_1n_5$ times around a circle; this circle is in  the direction
$x_5\equiv y$ and has length $2\pi
R$.  Among the various supergravity quanta that can be considered,
the simplest are the
s-wave ($l=0$)  modes of minimally coupled scalars. Let the compact
4-manifold $M$ be  $T^4$,
spanned by the coordinates $x_6, x_7, x_8, x_9$. Then the fluctuation
$h_{67}$ is an example of
such a minimally coupled scalar, and we will use it for illustration
in what follows.

When an incident quantum $h_{67}$ encounters the effective string,
its energy can get
converted to that of vibrations of the string \cite{cm, dm1, Wad96}.
At low energies, the
dominant process is the
absorption of the s-wave, and this creates one left moving vibration
quantum  and one right
moving vibration quantum  on the string. The incident quantum has an
energy $\omega$, and
its momentum lies only in the noncompact directions $x_1, x_2, x_3,
x_4$. The vibrations of the
string have momenta along $x_5$, and thus the energy momentum vectors
are respectively
            for the left moving quantum   and  the right
moving  quantum
\be
(p_0=\omega/2, p_5=-\omega/2) ~~{\rm and}~~(p_0=\omega/2, p_5=\omega/2)
\label{six}
\ee
            Further, one of
these vibrations   is polarized in the direction
$x_6$ and the other is polarized in the direction $x_7$. The graviton
in fact corresponds to the
symmetric combination
\be
h_{67}\rightarrow {1\over \sqrt {2}}~(|x_6\rangle_L\times
|x_7\rangle_R+|x_7\rangle_L\times
|x_6\rangle_R)
\label{seven}
\ee
(The antisymmetric combination corresponds to absorption of the
Ramond-Ramond 2-form
$B^{RR}_{67}$.)

The amplitude for this absorption was computed in \cite{dm1, dm2}
from the action for
the effective string coupled
to gravity. It was found that such a calculation
with the effective string
gave the same absorption cross section as that for absorption of low
energy s-wave minimally
coupled scalars into the D1-D5-momentum black hole.  In \cite{ms1}
it was noted that this
agreement persisted even if we do not have a large momentum charge,
and in fact was also true
to higher orders in the energy of the scalar as long as the D1 and D5
charges were kept large.
This indicates that the simplest system that captures the effective
physics would be the D1-D5
system with no momentum excitations; any momentum charge would be
considered one of the
possible excitations of our starting configuration. We will thus work
with states of the D1-D5
system with no net momentum charge in this paper.

\b

{\subsection{Time delay and its microscopic interpretation.}}

Let us first consider the state in the R sector that results from
spectral flow of the chiral
primaries
\be
\sigma_N^{s, \bar s}
\ee
In this case the twist operator joins together all $N=n_1n_5$ strands of the
effective string into one single
`multiply wound' strand, which was drawn schematically in figure
\ref{figGround}(a).

This state corresponds to the string configuration used in
\cite{msuss, dm1, dm2, ms1} in the
study of black hole entropy and absorption.
In our present considerations we will be also interested in the spin
of this effective string.
As discussed in the previous section, the variables $s, \bar s$ take
the values $\pm {1\over 2}$,
and  in the R sector the effective string forms  a multiplet with
\be
j_L=j_R={1\over 2}
\ee
\begin{figure}
\epsfxsize=5in \epsffile{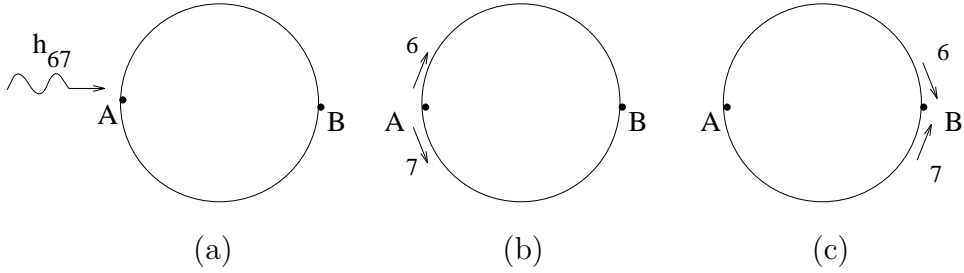}
\begin{picture}(350,30)
\put(70,10){(a)}
\put(187,10){(b)}
\put(304,10){(c)}
\end{picture}
\caption{(a) A graviton is incident on the component string; \ (b)
The graviton is converted to a pair
of vibration modes; \ (c) The vibration modes meet again at B.
\label{figHitString}}
\end{figure}
To study the picture of absorption by this effective string  we have
redrawn in figure
\ref{figHitString}  the string `opened
up' into a large circle; the length of this circle is $2\pi R
n_1n_5$.  The incident graviton $h_{67}$
breaks up into a pair of vibration modes at the point marked $A$.
Since the graviton had no
momentum along the direction $x_5\equiv y$, the amplitude for this process is
the same for any
position of the point $A$ along the string. The wavefunction of the
two vibration quanta is thus
of the form
\be\label{NoteMinus}
\Psi(y_1, y_2)=\psi(y_1-y_2)
\ee
Here  $y_1, y_2$ are the coordinates of the left moving and right
moving vibration
modes on the  effective string; thus $0\le y_i< 2\pi R n_1n_5$ in the
present example. (We can
think of these modes as massless open string states traveling along
the effective string, and
then
$y_1, y_2$ are the locations of the these open strings.) The modes
travel in opposite
directions at the speed of light $v=c=1$, and  encounter each other
again at the point
$B$ which is halfway along the effective string. The time interval
between the start at $A$ and
the meeting at
$B$ is
\be
\Delta t_{CFT} = {2\pi R n_1n_5\over 2}= \pi R n_1n_5
\label{toneCFT}
\ee
As we will see below, there is only a small chance that the modes
interact and re-emerge as a
graviton at the point $B$. If they do not interact to leave the
effective string, then they
travel around again and re-encounter each other at the point $A$,
after a further  time
(\ref{toneCFT}).

As mentioned already there is nothing special about the points $A, B$
on the string, since the
center of mass of the two vibration modes is uniformly smeared over
the string. The only
physical quantity therefore is the time between successive encounters
of the modes (not the
place where they interact), and this time is given by (\ref{toneCFT}).

Now let us look at the  supergravity background corresponding to this
R sector state. As
summarized in the last section, the D1-D5 geometry  with angular momentum
$j_L=j_R=j$ has a throat terminating  after a certain distance.   If
a quantum of  a minimally
coupled scalar is incident from infinity, then there is a small
probability that it enters the
throat. If it does enter the throat, then it travels to the end where
it reflects back and travels
again to the start of the throat. Since the probability to enter the
throat was small, the
probability to leave the throat is also small, since these two
probabilities are equal. Thus the
quantum travels several times up and down the throat before exiting
the throat back to spatial
infinity. The time for traveling once down the throat and back was
computed in \cite{lm4} and
is given by (eqn. (\ref{tthree}))
\be
\Delta t_{SUGRA}={\pi R\over \gamma}={\pi R n_1n_5\over 2j}=\pi n_1n_5 R
\label{timeSUGRA}
\ee
where we have used the fact that we have $j={1\over 2}$.

This time $\Delta t_{SUGRA}$ {\it exactly} equals the time $\Delta
t_{CFT}$ found in the
microscopic computation in (\ref{toneCFT}) above.

The essential idea of the AdS/CFT correspondence then yields the
following picture:

\b

(i)\q The effective string is dual    to the throat region of the
supergravity solution.
It is important to note here that the throat  is not infinitely long;
otherwise we could
not have found the above
relation between $\Delta t_{SUGRA}$ and $\Delta t_{CFT}$.

\b

(ii)\q  A graviton outside the throat in the supergravity solution is
described by the graviton
being present outside the effective string in the CFT picture.  Note
that we are
working with quanta of wavelength
$\lambda$ much larger than the scale  $(Q_1Q_5)^{1/4}$, so this
quantum outside the throat
effectively travels in a flat metric in the supergravity solution.

\b

(iii) \q The process where the quantum enters the throat in the
supergravity solution  maps to
the process in the CFT where the incoming graviton converts its
energy to that of the two
vibration modes of the effective string. Similarly the process where
the graviton manages to
leave the throat and escape to infinity maps to the process in the
CFT where the vibration
modes collide and leave the effective string as a single graviton.

\b

            (iv) \q  From (iii) above it is logical to   identify the
supergravity state where the quantum
is near the start of the throat with the CFT configuration where the
two vibration modes are
close to each other, as at points $A, B$ in figure \ref{figHitString}.
The quantum at
the end of the throat (near
$r=0$) maps to the CFT configuration where the vibration modes are separated by
the maximal possible  distance  along the effective string. This is
of course just a version of the
UV/IR correspondence \cite{pol}.

\b

(v)\q The supergravity quantum travels several times up and down the
throat, with a small
probability to escape each time it reaches the start of the throat.
Correspondingly, the
vibration modes travel around the string, with a small probability to
collide and leave the string
as a graviton each time they meet.

\b

In the following  we will make this correspondence more precise, and provide
additional evidence for the proposed relation between  CFT
excitations  and their description in
the supergravity dual.

\b

{\subsection{General values of $j$.}}

As we will see later in the section on black hole formation, for the
case $j_L=j_R={1\over 2}$
considered above the supergravity solution is not trustworthy all the
way to the end of the
throat. More precisely, the backreaction of the quantum placed in the
throat deforms
the throat in a significant way, at least near the end of the throat.
To avoid this we now look at
larger values of angular momentum, where it will turn out that the
backreaction of the
quantum can indeed be ignored.

\begin{figure}
\begin{tabular}{c}
\epsfxsize=3in \epsffile{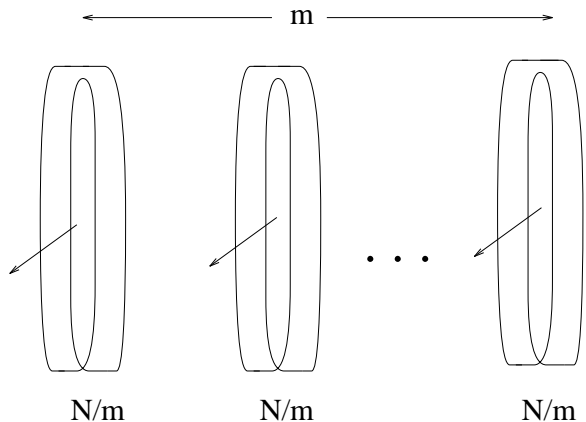}
\end{tabular}
\caption{Component strings for the state $[\sigma_{N/m}^{--}]^m$. The
spins are all
aligned to give $j=m/2$.\label{figCompString}}
\end{figure}

Let us look at a state of the CFT with
\be
j_L=j_R=j={m\over 2}
\label{five}
\ee
The simplest way to make such a state is to take the chiral primary
in the NS sector
\be
[\sigma_{N\over m}^{--}]^m
\ee
Upon spectral flow to the R sector, each component $\sigma_{N\over
m}^{--}$ gives a
`multiwound' string that is wound
\be
n={N\over m}={n_1n_5\over m}
\ee
times around the circle $x_5\equiv y$. There are $m$ such component strings.
We sketch this
configuration in figure \ref{figCompString}.  The spin of each multiwound
component is
$j_L=j_R={1\over 2}$, and in the state that we have taken all these
spins are aligned to produce
the spin values (\ref{five}).

An incident graviton can be absorbed into any of these $m$ component
strings. But now the
travel time for the vibration modes around the string, before they
meet each other again, is
\be
\Delta t_{CFT}={\pi n_1 n_5 R\over m}
\label{sthree}
\ee

Now we look at the corresponding supergravity solution. When the
angular momenta are given
by (\ref{five}) then the  time for a quantum to travel once down and back
up the throat is (eqn.(\ref{tthree}))
\be
\Delta t_{SUGRA}={\pi  R\over \gamma}={\pi R n_1n_5\over 2j}={\pi
n_1n_5 R\over m}
\label{tfour}
\ee
Thus again we find
\be
\Delta t_{CFT}=\Delta t_{SUGRA}
\label{tfive}
\ee

We have used the s-wave quanta as an illustration but we see
immediately that we get
the relation (\ref{tfive}) for higher partial waves as well. The
supergravity travel
time $\Delta t_{SUGRA}$ was found in \cite{lm4} to be independent of
the harmonic
order $l$. In the CFT the absorption of a higher partial wave
creates  set of massless left movers $\partial X \psi\dots \psi$ and a set of
massless right movers
$\bar\partial X \bar \psi\dots \bar\psi$ \cite{sm2, mang, gubser}. Each set of
excitations travels at the speed of light around the component
string,  so that
$\Delta t_{CFT}$ is independent of $l$ as well.

\b

\subsection{The rate of radiation.}

Suppose that we place an s-wave scalar quantum in the supergravity
throat; the dual CFT state has a pair of vibration modes on
the effective string. We wish to compute the probability per unit
time for the scalar quantum to escape from the throat, and
compare this to the probability per unit time for the vibration modes
on the effective string to collide and leave the string.

         First
we look at the CFT computation. Let there be one left
moving vibration and one right moving vibration on the string, with
the momentum vectors
(\ref{six}) and in the wavefunction corresponding to a graviton (\ref{seven}).
The interaction that leads to the emission of the graviton is given
by using the DBI action for the
effective string. The computation is essentially the same as that
done for   absorption into the
black hole microstate, but since we are here  looking at the emission
rate we reproduce the
relevant calculation in Appendix A. Consider the R ground state
arising from the chiral primary
$[\sigma_{N/m}^{--}]^m$. This state has $m$ component strings, all in
the same state, and
the spins are aligned to give   $j_l=j_R={m\over 2}$.
Then the  probability
per unit time for the
vibrations to collide and emit a graviton is found to be
\be
{\cal R}_{CFT}=m{\pi^2\omega^4 g^2\over 2 V (2\pi) R}
\ee
Here $g$ is the string coupling, $(2\pi)^4 V$ is the volume of the
$T^4$ and we have set
$\alpha'=1$.

   Now we look at the supergravity solution. The throat is assumed to be
long compared to the
scale $(Q_1Q_5)^{1/4}$ of the geometry near the start of the throat.
A quantum incident from
infinity has a probability ${ P}$ to enter the throat, after
which it propagates down the
throat till it reaches the end where it reflects. The long
length of the throat implies
that the probability ${ P}$ depends only on the geometry near
the start of the throat, which
is the same as the geometry in the nonrotating case $j=0$. Thus we
can read this probability off
from \cite{ms1, lm4}, and we find (using eqns (\ref{sone}), (\ref{mafour}))
\be
{ P}={\pi^2 \over 4}Q_1Q_5\omega^4={\pi^2\over 4} \omega^4
{n_1n_5 g^2\over V}
\label{pone}
\ee
The number of times the quantum tries to exit the throat per unit time is
\be
{1\over \Delta t_{SUGRA}}
\ee
Thus the probability of emission per unit time is (using
(\ref{tfour}), (\ref{pone}))
\be
{\cal R}_{SUGRA}={{ P}\over \Delta t_{SUGRA}}=m{\pi^2\over 2}
\omega^4 { g^2\over V} {1\over 2\pi R}
\ee

We see that
\be
{\cal R}_{CFT}={\cal R}_{SUGRA}
\label{eight}
\ee
so that the rates of emission agree {\it exactly}
between the supergravity and CFT pictures.

Let us compare the above calculation with  the computations
\cite{dm1, dm2, ms1} that show
agreement between
microscopic radiation rates and Hawking radiation rates from the
corresponding black holes. The
CFT computation is the same in each case. But on the supergravity
side  in the present case we do not have a
horizon, and the graviton does not represent Hawking radiation. We
are in fact below the black
hole formation threshold, and so instead of finding a horizon
somewhere down the throat we
find a point from which we can reflect back. The travel time $\Delta t_{SUGRA}$
has no analogue if we have a horizon. Thus the agreement
(\ref{eight}), while similar to that in
the black hole case, has a somewhat different interpretation.

\b

\section{Generic microstates and the dual throat geometries}
\renewcommand{\theequation}{4.\arabic{equation}}
\setcounter{equation}{0}

So far we have used specific examples of D1-D5 microstates to argue
that  $\Delta t_{CFT}=\Delta t_{SUGRA}$.  For this to
be a general principle,
however, we must examine all possible R ground states of the D1-D5 system.

\subsection{Unbound states: a potential difficulty and its resolution}

A simple way to violate $\Delta t_{CFT}=\Delta t_{SUGRA}$ would
appear to be the following. Take
two D1-D5 bound states, with angular momenta
\be
(j_L=j^3_L={m\over 2}; ~j_R=j^3_R={m\over 2}) ~~{\rm and }
~~(j_L=-j^3_L={m\over 2};
~j_R=-j^3_R={m\over 2})
\ee
Since such D1-D5 states are mutually BPS, we can construct a
supergravity solution with
arbitrary locations for the centers of the two parts.  Let the
two centers be coincident. The
angular momentum of the two bound states cancel each other, so that
we get $j_L=j_R=0$.
Naively, looking at eqn. (\ref{tone}),  (\ref{ttwo}) one  may think that this
geometry will have an infinite
throat and infinite $\Delta
t_{SUGRA}$. But looking at the lengths of the component strings in
the CFT we still find $\Delta
t_{CFT}={\pi n_1 n_5 R\over m}$, so that we have a
potential contradiction.

But a closer look at the supergravity solution reveals the following.
The rotation parameter $a$ in
the solution (\ref{MaldToCompare}) multiplies $dtd\phi$, but it also
appears in the
function $f=r^2+a^2\cos^2\theta$.  Suppose we take a solution
(\ref{MaldToCompare}) with rotation
parameter $a$ and superpose another solution with the same charges and rotation
parameter ${\tilde a}=-a$, and the same center. Then the coefficient of
$dtd\phi$ disappears, but $f$
remains unchanged since it does not depend on the sign of $a$.  We
have developed in
Appendix \ref{AppGener} an extension of the chiral null models to yield
directly general solutions for
the D1-D5 system. The 6-D Einstein metric for this superposition,
            worked out in Appendix \ref{AppUnbound}, is
\bea\label{myToComparep}
ds_E^2&=&-\frac{{ f}_0}{\sqrt{{ f}_1
{ f}_{5}}}(dt^2-d{ y}^2)
+\sqrt{{ f}_1 { f}_{5}}(\frac{dr^2}{r^2+a^2}+d\theta^2)
\nonumber\\
&+&\frac{\sqrt{{ f}_1 { f}_{5}}}{{ f}_0}
\left[r^2\cos^2\theta d\psi^2+
(r^2+a^2)\sin^2\theta d\phi^2\right]
\eea
where
\be
f_0=r^2+a^2\cos^2\theta,~~ f_1=f_0+Q_1, ~~~f_5=f_0+Q_5
\ee

We cannot solve the wave equation in this metric  since the variables do
not separate. But we can
estimate the time delay by looking at geodesics in the throat. A
generic geodesic goes down the
throat only down to
$r\sim a$; the travel time is then\footnote{There exist an
exceptional set of geodesics that head directly into the
singularity  at $r=0, \theta=\pi/2$, which do not return back in a
finite time $t$. As we will note below (and this is
explained in more detail in Appendix \ref{AppSelfInt}) this situation is not
generic; in a generic solution the geodesics (and
waves) reflect back in  a finite time from the singularity. We also
argue in Appendix \ref{AppSelfInt} that quantum
fluctuations of the metric will result in  $\Delta
t_{exceptional}\sim \Delta t_{CFT} \times \log j$ even for such
nongeneric geometries.}
\be
\Delta t_{SUGRA}\sim \Delta t_{CFT}
\ee

We can superpose several D1-D5 bound states, each having rotation but
such that the combined
angular momentum is zero. It is easy to generalize the above
construction to see that in each
case we still get $\Delta t_{SUGRA}\sim \Delta t_{CFT}$.

\subsection{ Arbitrary $j$ configurations of the bound state}

Let us now return to our principal consideration -- the analysis of a
single D1-D5 bound
state. Here again we face the following
potential problem.  Consider the R ground state that results from a
chiral primary
made of $m$ components, each of the form
$\sigma^{s,\bar s}_{N/m}$. If the spins $s,\bar s$ are all aligned then
we get $j_L=j_R={m\over 2}$; this is the
maximum value of $j$ in
(\ref{nine}) and gives the configuration we studied in the  section 
\ref{secthree}. But
from (\ref{nine}) we see that we can also combine the spins $s, \bar 
s$ to obtain states
with lower $j$, all the way
down to $j\approx 0$.  Does the effective length of the throat depend
on the value of $m$ or on
the value of $j$? If it depends on $j$ then we have a problem; by
choosing $m\gg 1$ we make the
component strings of the CFT small (length $\sim 2\pi R N/m$ each) but the
supergravity throat
could be made long by letting $j\approx 0$. We will see
however that the supergravity throat terminates at a distance that
depends on $m$ rather
than $j$. This fact will be crucial to our conjecture that $\Delta
t_{CFT}= \Delta t_{SUGRA}$.

\subsubsection{The duality map from D1-D5 to FP}

To see the nature of supergravity solutions corresponding to general
values of $j$ in
(\ref{nine}) we will map the D1-D5 system by a chain of
dualities to the FP system:
F represents fundamental string winding  (along $x_5\equiv y$) and P
represents momentum
            (also along
$y$). The reason for starting the analysis with the FP system is
the following.
The $M^N/S^N$ orbifold CFT describes a single D1-D5 bound state. To
study the supergravity dual we need the
metric for a single bound state system,  and not for a superposition
of many such bound states. In the FP language a
single bound state is easy to characterize: it  arises from a single
multiwound string carrying left moving
vibrations. Duality then maps such FP solutions to metrics of single
D1-D5 bound states.

The sequence of dualities needed to map the D1-D5 system to the FP system
is easy to write, and can be found for example in \cite{lm3}. The following
is the relation of quantities and states in the two systems:

\b

(i)\q The number of D5-branes in the D1-D5 system becomes the winding number
$n_w$ of the fundamental string (i.e. $n_w=n_5$). The number of D1-branes
becomes the number of units of momentum ($n_p=n_1$). The bound state of the
D1-D5 system maps to a bound state of the winding and momentum modes of the
FP system. Let the radius of the $y$ circle after dualities be $R'$. The
string of the FP system then has a total length $L_T=2\pi R' n_w$. Momentum
modes on this string come in fractional units
\be
p={2\pi n\over L_T}={n\over n_w R'}
\ee
with the constraint that the total momentum has the form
\be
P_T={n_p\over R'}={n_wn_p\over n_w R'}
\ee
with $n_p$ an integer \cite{dm}. Thus on the fundamental string of total length
$L_T$ the total level number of excitations is
\be
N=n_pn_w=n_1n_5
\ee

\b

(ii)\q Let the fundamental string carry a quantum of vibration in a
transverse direction
$ i $, $i=1,2,3,4$ with wavelength $\lambda={L_T/n}$. This excitation is
generated by the oscillator $\alpha^i_{-n}$. If the string has
$n_p=N/n_w$ units of momentum then the total state of the string
is of the form
\be
[\alpha^{i_1}_{-n_1}]^{m_1}\dots[\alpha^{i_k}_{-n_k}]^{m_k}|0\rangle
\label{qone}
\ee
with
\be
\sum_j m_j n_j = N
\ee

The state (\ref{qone}) corresponds in the D1-D5 system to the chiral primary
\be
[\sigma_{n_1}^{s_1, \bar s_1}]^{m_1}\dots[\sigma_{n_k}^{s_k, \bar s_k}]^{m_k}
\ee
Thus  in the R ground state of the D1-D5 system if we have a
component string  that is
      wound
$n$ times around
$y$ then this component string  maps to a momentum mode
$\alpha_{-n}$, which has
wavelength
$L_T/n$ around the fundamental string in the FP system. The component
string in the D1-D5
system has four polarization states $(s, \bar s)=(\pm{1\over 2},
\pm{1\over 2})$ which gives the
spins under $SU(2)_L\times SU(2)_R=SO(4)$, and these map to the four
polarizations $\alpha^i$
of the momentum mode.\footnote{The $M^N/S^N$ orbifold does not
include the center of mass
U(1) factor of the D1-D5 system, so we see only one vacuum
$|0\rangle$ in (\ref{qone}) rather
than all the ground states of a single fundamental string without
momentum. Further we are
concentrating on only those chiral primaries in the D1-D5 system that
arise from the $(0,0),
(0,2), (2,0), (2,2)$ forms on the 4-manifold $M$, so we  write only
the polarizations $\alpha^i$
for the momentum modes rather than all possible polarizations.}

\b

\subsubsection{ Mapping D1-D5 states to FP configurations}

Let us study the FP representation of some selected  R ground states
of the D1-D5 system.

\b

\begin{figure}
\begin{tabular}{cc}
\begin{tabular}{c}
\begin{picture}(200.00,0.00)(0,-40)
\begin{picture}(50.00,50.00)(0.00,0.00)
%
%
\qbezier(15,15)(20,15)(20,35)
\qbezier(20,35)(20,45)(18,45)
\qbezier(18,45)(16,45)(16,35)
\qbezier(16,35)(16,15)(21,15)
\qbezier(15,15)(12,15)(12,35)
\qbezier(21,15)(24,15)(24,35)
\qbezier(12,35)(12,50)(18,50)
\qbezier(24,35)(24,50)(18,50)
\put(18,50){\line(1,0){0}}
\put(5,5){$n=2$}
\put(19,32){\vector(-2,-1){20}}
\end{picture}
\begin{picture}(50.00,50.00)(10.00,0.00)
%
%
\qbezier(15,15)(20,15)(20,35)
\qbezier(20,35)(20,45)(18,45)
\qbezier(18,45)(16,45)(16,35)
\qbezier(16,35)(16,15)(21,15)
\qbezier(15,15)(12,15)(12,35)
\qbezier(21,15)(24,15)(24,35)
\qbezier(12,35)(12,50)(18,50)
\qbezier(24,35)(24,50)(18,50)
\put(18,50){\line(1,0){0}}
\put(5,5){$n=2$}
\put(19,32){\vector(-2,-1){20}}
\end{picture}
\put(0,25){\circle*{2}}
\put(10,25){\circle*{2}}
\put(20,25){\circle*{2}}
\begin{picture}(50.00,50.00)(-50.00,0.00)
%
%
\qbezier(15,15)(20,15)(20,35)
\qbezier(20,35)(20,45)(18,45)
\qbezier(18,45)(16,45)(16,35)
\qbezier(16,35)(16,15)(21,15)
\qbezier(15,15)(12,15)(12,35)
\qbezier(21,15)(24,15)(24,35)
\qbezier(12,35)(12,50)(18,50)
\qbezier(24,35)(24,50)(18,50)
\put(18,50){\line(1,0){0}}
\put(5,5){$n=2$}
\put(19,32){\vector(-2,-1){20}}
\end{picture}
\begin{picture}(20.00,50.00)(10.00,-15.00)
\put(50,14){$2\pi R$}
\put(60,25){\vector(0,1){10}}
\put(60,10){\vector(0,-1){10}}
\end{picture}
\put(-60,60){\vector(1,0){60}}
\put(-90,60){\vector(-1,0){70}}
\put(-85,57){$N/2$}
\put(-85,-30){(a)}
\end{picture}
\\
\bigskip\\
\epsfxsize=2in \epsffile{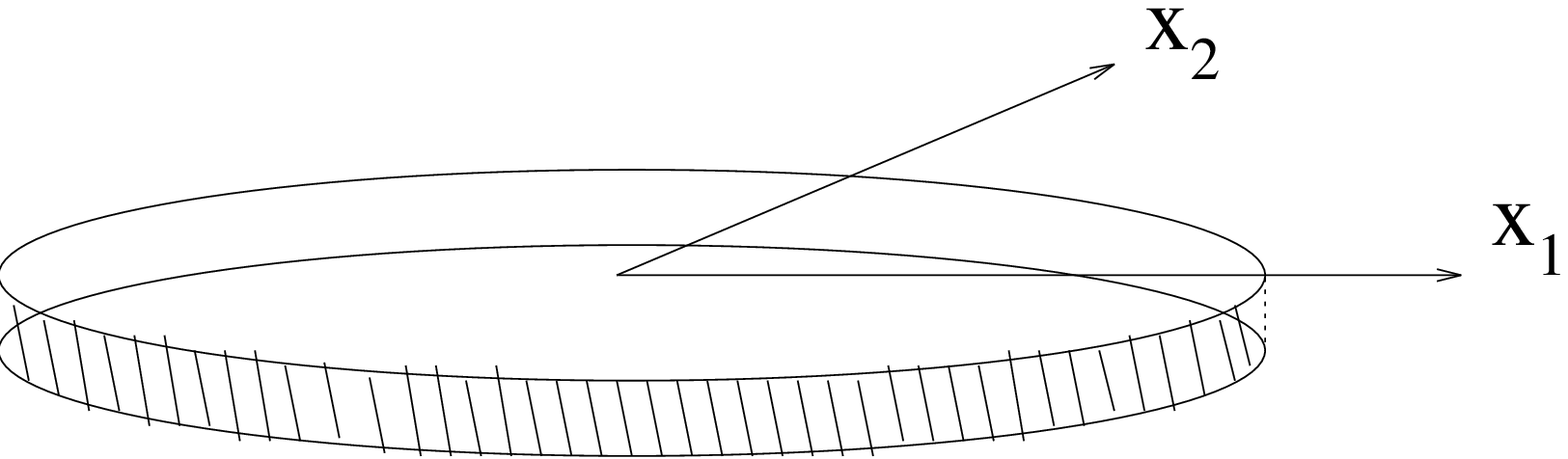}\\
(c)\\
\bigskip\bigskip\\
\epsfxsize=2in \epsffile{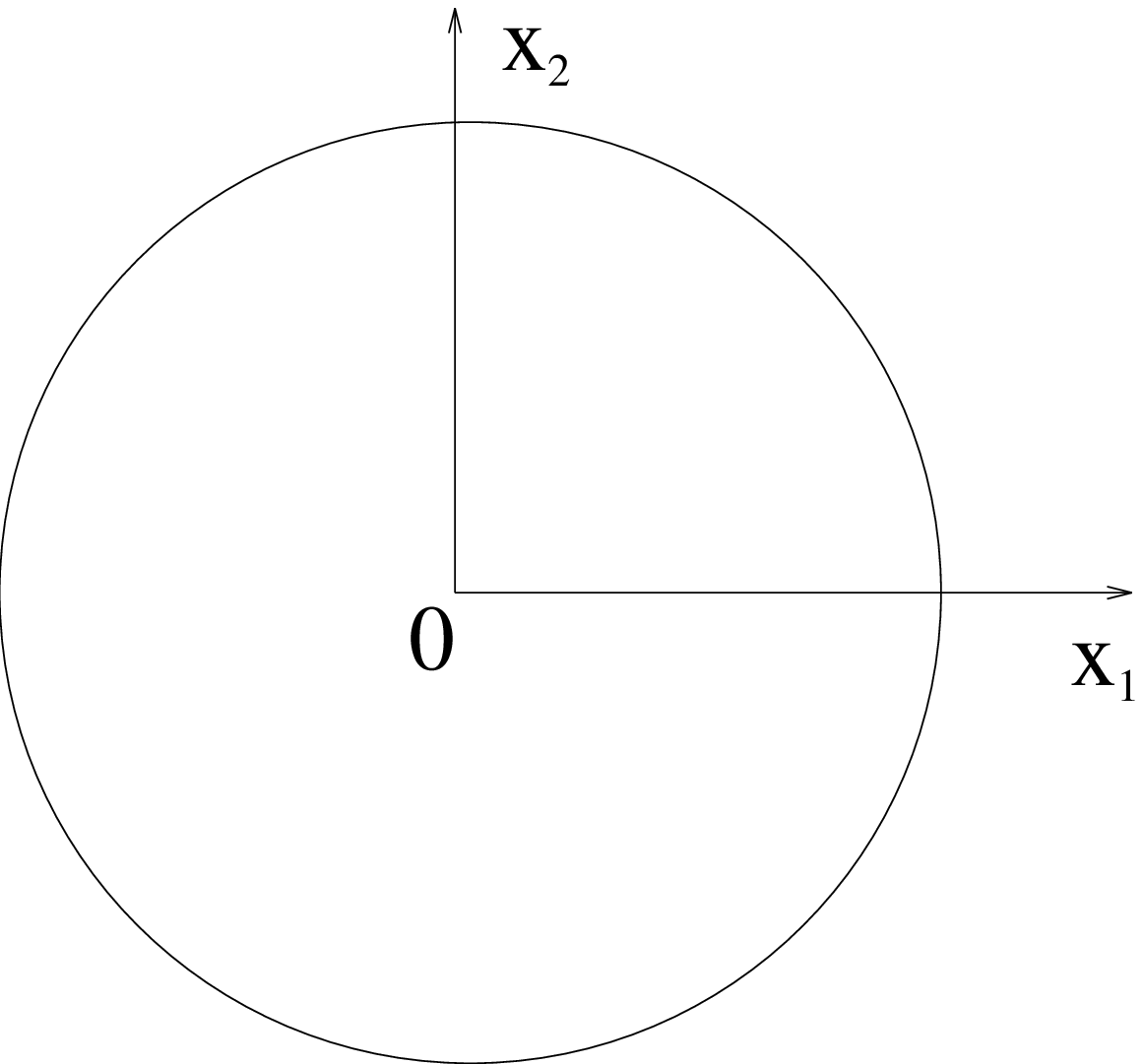}\\
(d)
\end{tabular}
&
\begin{picture}(20.00,300.00)(-90.00,140.00)
%
%
%
\qbezier(20,15)(40,30)(40,60)
\qbezier(40,60)(40,90)(20,105)
\qbezier(20,105)(0,120)(0,150)
\qbezier(0,150)(0,180)(20,195)
\qbezier(20,195)(40,210)(40,240)
\qbezier(40,240)(40,270)(20,285)
\qbezier(20,285)(0,300)(0,330)
\qbezier(0,330)(0,360)(20,375)
%
%
\put(20,375){\line(1,0){55}}
\put(20,15){\line(1,0){55}}
\put(75,180){\vector(0,-1){165}}
\put(75,205){\vector(0,1){170}}
\put(55,190){$2\pi n_w R'$}
%
\put(0,327){\line(1,0){25}}
\put(0,333){\line(1,0){25}}
\put(30,327){$2\pi R'$}
\put(23,319){\vector(0,1){8}}
\put(23,341){\vector(0,-1){8}}
\put(20,0){(b)}
\end{picture}
\end{tabular}
\\
\caption{
(a)  Component strings for the D1-D5 microstate 
$[\sigma_2^{--}]^{N/2}$. \q (b) The fundamental string
in the dual FP system, oscillating in the harmonic $n=2$, shown 
opened up to its full length $2\pi n_w R'$.
\q (c) The string of (b) as it actually appears due to the 
identification $y'\rightarrow y'+2\pi R'$. \q (d)
The cross section of the singularity created by the strands in (c) in 
the classical limit.
\label{figCirStr}
}
\end{figure}

(i)\q Take the chiral primary
\be
[\sigma_n^{--}]^{N/n}
\label{qtwo}
\ee
of the NS sector of the D1-D5 system and consider its corresponding R
ground state.  There are $N/n$
component strings of
winding number $n$ each.   Figure \ref{figCirStr}(a) depicts this
state for $n=2$.

%


In
the FP system we have
$N/n$ quanta of the
$n$th harmonic around the fundamental string. Further, since the spins
of the factors in
(\ref{qtwo}) are all aligned, the angular momentum is
$j_L=j_R={N\over 2n}$. The fundamental
string rotates in a circle; the metric of this state and the duality maps
were constructed in
\cite{lm3}.  Let the rotation generators of the two $SU(2)$ factors and the
$SO(4)$ be related such that
$j^3_L={1\over 2} (M_{12}+M_{34}), j^3_R={1\over 2} (M_{12}-M_{34})$;
then the choice $(j^3_L,
j^3_R)=(-{1\over 2}, -{1\over 2})$ for each $\sigma_n$ in (\ref{qtwo})
implies that the rotation of
the string is in the
$x_1$-$x_2$ plane.

This F string is pictured in figure \ref{figCirStr}(b), opened up to its total
length $2\pi R'
n_w=2\pi R' n_5$. Due to the identification $y\rightarrow y+2\pi R'$ the
string actually covers a cylindrical surface in a helical fashion;
this is pictured in figure \ref{figCirStr}(c).
We can characterize this cylinder by its cross section, which is a
circle, depicted in figure \ref{figCirStr}(d).

Mapping by dualities back to the D1-D5 system we get the metrics
(\ref{MaldToCompare}) with
$j_L=j_R={N\over 2n}$.
If we take $n=1$ then we get the maximal possible
angular momentum $(j_L,
j_R)=({N\over 2}, {N\over 2})$; it was shown in \cite{mm} that
the corresponding
D1-D5 supergravity solution was a smooth geometry.

\b

\begin{figure}
\begin{tabular}{cc}
\begin{picture}(200.00,20.00)
\begin{picture}(50.00,150.00)(0.00,0.00)
%
%
\qbezier(15,15)(20,15)(20,35)
\qbezier(20,35)(20,45)(18,45)
\qbezier(18,45)(16,45)(16,35)
\qbezier(16,35)(16,15)(21,15)
\qbezier(15,15)(12,15)(12,35)
\qbezier(21,15)(24,15)(24,35)
\qbezier(12,35)(12,50)(18,50)
\qbezier(24,35)(24,50)(18,50)
\put(18,50){\line(1,0){0}}
\put(15,5){$n$}
\put(19,32){\vector(-2,-1){20}}
\end{picture}
\begin{picture}(50.00,150.00)(15.00,0.00)
%
%
\qbezier(15,15)(20,15)(20,35)
\qbezier(20,35)(20,45)(18,45)
\qbezier(18,45)(16,45)(16,35)
\qbezier(16,35)(16,15)(21,15)
\qbezier(15,15)(12,15)(12,35)
\qbezier(21,15)(24,15)(24,35)
\qbezier(12,35)(12,50)(18,50)
\qbezier(24,35)(24,50)(18,50)
\put(18,50){\line(1,0){0}}
\put(15,5){$n$}
\put(19,32){\vector(2,1){20}}
\end{picture}
\put(-10,25){\circle*{2}}
\put(0,25){\circle*{2}}
\put(10,25){\circle*{2}}
\begin{picture}(50.00,150.00)(-30.00,0.00)
%
%
\qbezier(15,15)(20,15)(20,35)
\qbezier(20,35)(20,45)(18,45)
\qbezier(18,45)(16,45)(16,35)
\qbezier(16,35)(16,15)(21,15)
\qbezier(15,15)(12,15)(12,35)
\qbezier(21,15)(24,15)(24,35)
\qbezier(12,35)(12,50)(18,50)
\qbezier(24,35)(24,50)(18,50)
\put(18,50){\line(1,0){0}}
\put(15,5){$n$}
\put(19,32){\vector(-2,-1){20}}
\end{picture}
\begin{picture}(20.00,150.00)(10.00,-15.00)
\end{picture}
\put(-70,60){\vector(1,0){45}}
\put(-110,60){\vector(-1,0){50}}
\put(-100,57){$N/n$}
\end{picture}
&
\epsfxsize=2.5in \epsffile{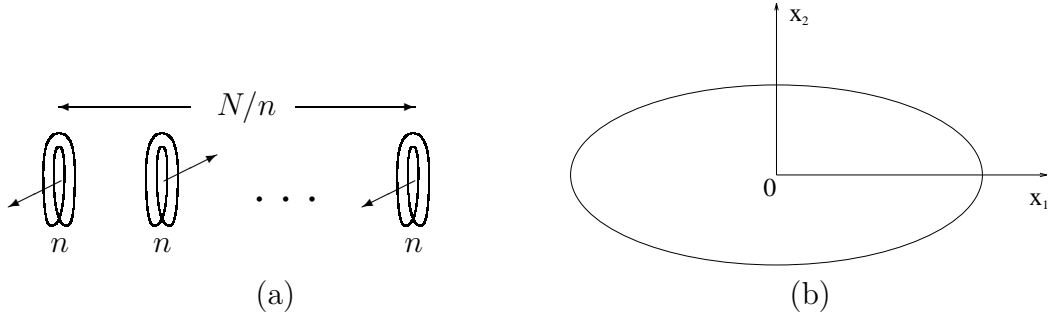}\\
(a)&(b)
\end{tabular}

\caption{(a) Component strings with spins {\it not} all aligned; ~
(b) The
shape of the singularity.\label{figEllStr}}
\end{figure}

(ii) \q The states described in (i)  correspond, in the list (\ref{nine}),
to $j$ taking its maximal
possible value ${N\over 2n}$.  If all the ${N\over n}$ spins are not aligned,
then from the same set of
`component strings' we can get smaller values of $j$. In the dual FP
system these latter states are
just obtained by taking the same harmonic on the fundamental string,
but making the string
swing around an ellipse rather than a circle:
\be
x'_1=A \cos( {n\over R' n_w} (y'+t')), ~~x'_2=B\sin ({n\over R' n_w}(y'+t'))
\label{qthree}
\ee
The string still forms a helix but covers a surface with
elliptical cross section as
depicted in figure \ref{figEllStr}. In particular letting the string
vibrate only in one direction (e.g.
$B=0$ in (\ref{qthree}))  would
give
$j_L=j_R=0$. The supergravity solution for this case is written down
explicitly in Appendix \ref{AppSelfInt}.

\b

\begin{figure}
\begin{tabular}{ccc}
\epsfxsize=1.5in \epsffile{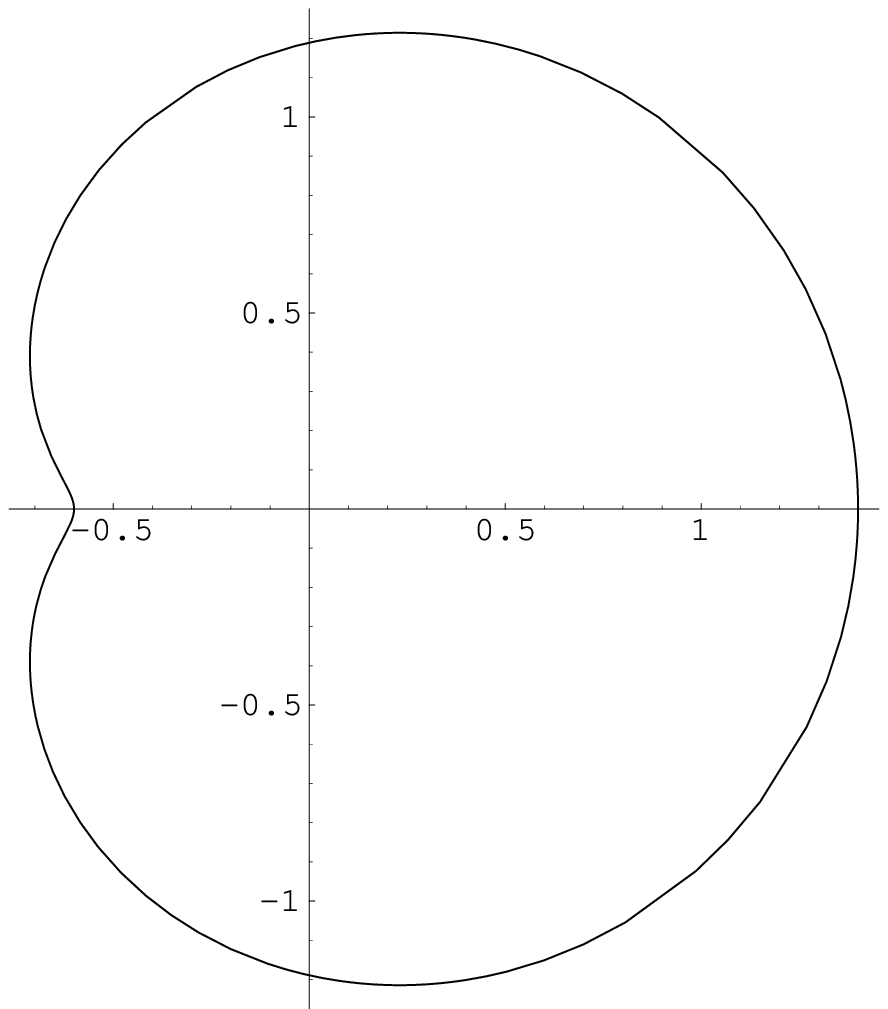}&
\begin{picture}(80.00,10.00)
\end{picture}
&
\epsfxsize=1.5in \epsffile{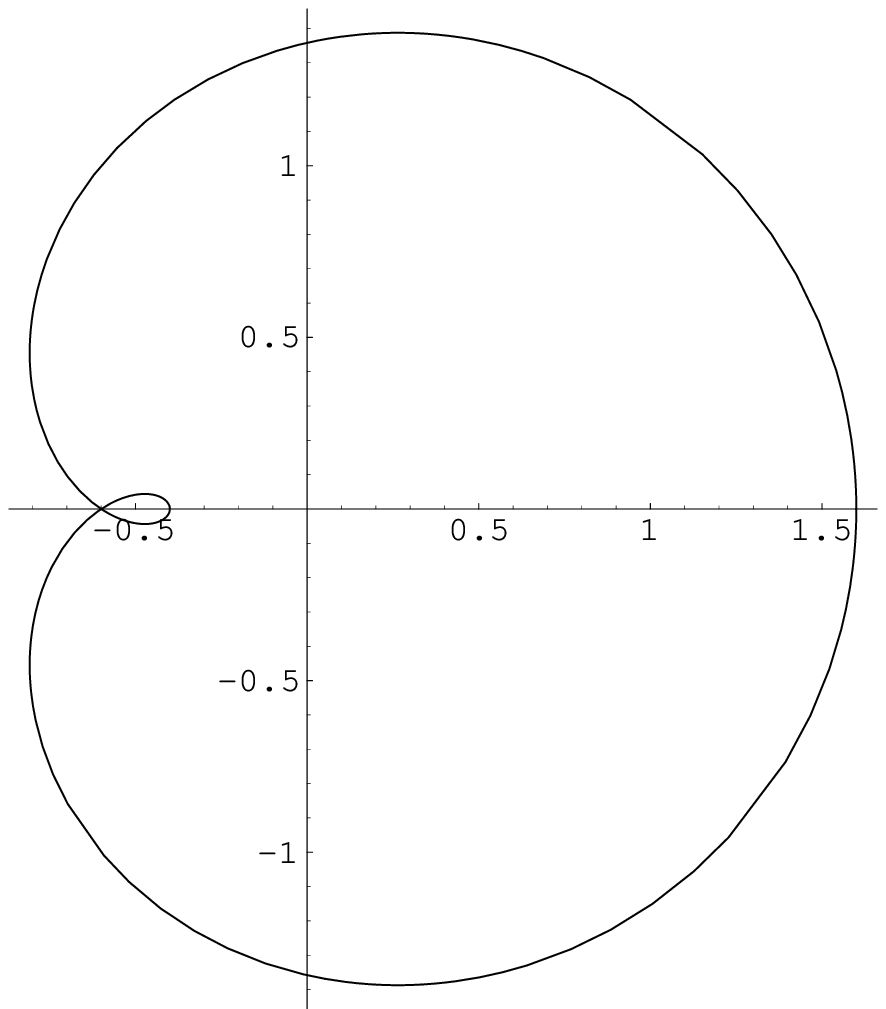}\\
(a)&&(b)
\end{tabular}
\caption{The singular curve when the FP string carries two different harmonics:
eqn. (\ref{TwoModesEqn}) parameter values (a) $A=1,B=0.4$; (b)
$A=1,B=0.6$.\label{figTwoStr}}
\end{figure}

(iii)\q Consider  a chiral primary that has twist operators of two
different orders, for example
\be
[\sigma_n^{--}]^{m_1}[\sigma_{2n}^{--}]^{m_2}, ~~~~~~nm_1+2nm_2=N
\ee
In the FP dual we have harmonics of order $n$ and $2n$ on the
string, so the waveform
looks like
\bea\label{TwoModesEqn}
x'_1&=&A_1 \cos ({n\over R'n_w}(y'+t')) + A_2 \cos (2{n\over R' n_w} (y'+t')),
\nonumber \\
x'_2&=&B_1
\sin ({n\over R' n_w}(y'+t')) + B_2
\sin (2{n\over R' n_w} (y'+t'))
\eea
The surface covered by the string now has a more complicated cross
section, depicted in figure \ref{figTwoStr}. For a range of
parameters $A_i, B_i$ the singularity
curve exhibits a self intersection. Such self-intersections are not
generically present
however, since the singular curve lies in a 4-dimensional space $\vec x'$.
\b

(iv) \q More generally, the singularity can have the shape of a
general curve. Let the string in the FP solution be described by the
profile $\vec x'=\vec G(v)$.
(Here $\vec x'$ is a 4 component vector $\{x'_1, x'_2, x'_3,
x'_4\}$.)  For a generic state the typical wavelength involved in the
oscillations is much larger than the compactification radius $R'$ of
the $y'$ coordinate.  Then for any fixed
$y'=y'_0$ the string passes through a set of points $\vec x'$ that
are closely spaced along a smooth curve. This
curve is given parametrically by
\be
\vec x'= \vec G (\alpha), ~~~0\le \alpha < 2\pi n_w R'.
\label{tYYfour}
\ee
           (By construction this curve is
independent of the choice of $y'_0$.)   In the classical limit the
spacing of points along the curve goes to zero.
Thus the classical geometry
that corresponds to this microscopic FP solution has a
singularity along the curve $\vec x'= \vec G (\alpha)$, with the
singularity extending over all $0\le y'<2\pi R'$ and
for all $t$. We will not explicitly mention in what follows the
extent in the $y', t$
directions, and characterize the singularity by the
shape of the singular curve  (\ref{tYYfour}).

It is important that in the classical limit the large value of the winding
$n_w$ gives a singularity structure that is invariant under
translations in $y'$. This invariance in $y'$ allows us to
apply  T-duality along $y'$, which is needed to obtain the
corresponding D1-D5 solution. The latter solution will also
have a singularity that is a curve in the space $\vec x$, and which
extends uniformly in the $y,t$ directions. The
singular curve for the D1-D5 geometry is
\be
\vec x=\vec F(\alpha), ~~~\vec F={g\over R\sqrt{V}}\vec G
\label{tYYfive}
\ee

Given that the curve (\ref{tYYfive}) is a 1-D subspace in 4 dimensions,
the following situation is generic:

\b

(i) \q $\vec x= \vec F (\alpha)$ is a simple closed curve with no
self-intersections.

\b

(ii)\q $|{\dot{\vec F}}(\alpha)|\ne 0$ at all points along the curve.

\b

Note that in the FP system if we take any  $y'=y'_0$  and look for
the points in the $\vec x'$ space through which the
string passes, then the number of these points per unit length along
the curve (\ref{tYYfour}) is
\be\label{PoinDens}
|{\dot{\vec G}}(\alpha)|^{-1}{1\over 2\pi R'}
\ee
The singularity is characterized by both its shape and the `density'
(\ref{PoinDens}).

Let us look at the D1-D5 system and examine the  geometry for a
general singular
curve $\vec F$.
By translational invariance we can set
\be \int_0^L dv F_i(v)=0. \ee
             We will also assume that singularity
in confined in the region with a typical size $a$, which means
that
\be F_i(v)F_i(v)\le a^2 \qquad\mbox{for all}\qquad v \ee

The throat geometry then has the following two properties:

\b

(i)\q
For ${\vec x}^2\gg a^2$ we get the metric (Appendix \ref{AppThroat})
\be\label{AsymSol1p}
ds^2\approx
-\frac{1}{h}(dt^2-dy^2)+hd{\vec x}d{\vec x},
\ee
where
\be\label{AsymSol2p}
h=\left[\left(1+\frac{Q_5}{{\vec x}^2}\right)
\left(1+\frac{Q_1}{{\vec x}^2}\right)\right]^{1/2}
\ee

If we compute the time of flight for a quantum from the start of the
throat to $r=a$ and back
then we get
\be
\Delta t_{SUGRA}=\pi {\sqrt{Q_1Q_5}\over a}
\label{sfour}
\ee

\b

(ii)\q The geometry ends smoothly at $r\sim a$ except for the
singular curve. We show in Appendix \ref{AppWave} that
waves incident on the singular curve in fact reflect from the curve,
as long as we have the genericity conditions (i)
and (ii) mentioned above for the curve. In Appendix \ref{AppGeodes} we
estimate the maximum time
$\Delta t_{sing}$ that a quantum can spend in the vicinity of the singularity
even if it heads radially into the singularity from $r\sim a$. We find that
\be
\Delta t_{sing}\sim {2\over \pi\sqrt{\bar n}} \Delta t_{SUGRA}
\ee
where $\bar n$ is the average winding number of the component
strings, and so $\bar n\ge
1$.\footnote{
Note that for the metrics (\ref{MaldToCompare}) we have performed
the exact computation for the travel time $\Delta t_{SUGRA}$ (eq.
(\ref{timeSUGRA}));  this time
includes all effects of approaching the singularity and returning
back, so that we do not need
to consider separately the time $\Delta t_{sing}$ spent near the singularity.}

\b

Putting together properties (i) and (ii) we find that
\be
\Delta t_{CFT}\sim \Delta t_{SUGRA}
\ee
We in fact expect an exact match between the CFT and  supergravity,
      but if all the component strings in
the CFT are not of equal length then $\Delta t_{CFT}$ can only be
defined upto some spread. Similarly the waveform
in the supergravity throat returns with some distortion and thus
$\Delta t_{SUGRA}$ has  a spread as well. For this
reason the above relation appears as an approximate one, but in the
cases where these time scales could be exactly
defined and computed we obtained the exact equality (\ref{tfive}).

\b

{\subsection{`Hair' in the D1-D5 geometry.}}

For geometries that have a black hole horizon there exists  the well
known  `no hair' conjecture,
which states that the microstates of the hole will not be visible in
its geometry. While this
conjecture may indeed not be valid as a general theorem, it is
nevertheless true  that we do not
know how to see the entropy of black holes as a count of different
spacetime geometries. Thus
for the D1-D5 momentum black hole in 4+1 dimensions, the metric
lifted to 6-D is
\be
\label{qfour}
d{s}_E^2=-\frac{1}{h}(d{t}^2-d{y}^2)+\frac{Q_P}{hr^2}(d{t}+d{y})^2
+h\left[dr^2+r^2d\Omega_3^2\right].
\ee
Classically this metric represents all the $\sim e^{2\pi
\sqrt{n_1n_5n_p}}$ microstates of
the hole.  If we set the
momentum charge to zero, we still find $\sim e^{2\sqrt{2}\pi
\sqrt{n_1n_5}}$ microstates. The
geometry (\ref{qfour}) reduces to
\bea\label{qfive}
d{s}_E^2=-\frac{1}{h}(d{t}^2-d{y}^2)+
h\left[dr^2+r^2d\Omega_3^2\right].
\eea
The classical horizon area is now zero, but the classical geometry
(\ref{qfive}) still seems to
exhibit a version of `no hair', since it is the same for all
microstates. How do we reconcile this
situation with the fact that we have found different throat
geometries for different microstates
of the D1-D5 system?

\begin{figure}
\epsfxsize=5in \epsffile{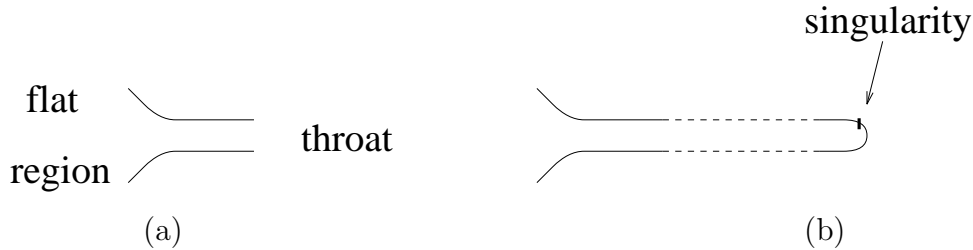}
\begin{picture}(350,20)
\put(50,0){(a)}
\put(300,0){(b)}
\end{picture}
\caption{\label{figThroat}(a) A leading order classical analysis
suggests an infinite throat; ~
(b) The throat actually ends at a distance that diverges as
$\hbar\rightarrow 0$.}
\end{figure}
It turns out that the geometry produced by generic microstates all
behave as (\ref{qfive}) upto
a `classical' distance down the throat (so that we see no hair) (figure
\ref{figThroat}a), but
start to differ at a distance
which is $\sim 1/\hbar$ down the throat, where we find the
the end of the throat and the above discussed singularities
(figure \ref{figThroat}b). More
precisely assume fixed the
parameters of the string background, and take the limit $n_1,
n_5\rightarrow \infty$ -- this is
the classical limit of the solutions. The radius of the throat is
$ (Q_1Q_5)^{1\over 4} \sim
(n_1n_5)^{1\over 4} $. Choose any dimensionless number $\mu$, and
look at distances  that are reached by a particle falling down the
throat for a time
       $t\sim \mu
(Q_1Q_5)^{1\over 4}$.  As
$n_1, n_5\rightarrow \infty$, the metrics for generic microstates
will all become the same in
this domain in the throat. The end of the throat comes for
$t\sim (n_1n_5)^{1/2}$, and in the classical limit this will appear to be
infinitely far down the throat. Thus we
recover a `no hair' classical limit, but see different geometries for
different microstates near the end of the throat.

\b

\subsection{A flat space computation to understand the size of the
singularity}

An essential part of the above results can be understood  by
considering the string of the FP system in
flat space. It was crucial in obtaining $\Delta t_{CFT}\sim \Delta
t_{SUGRA}$ that the size of the singularity, for fixed winding and
momentum charges, depended on the {\it wavelength} of the
oscillations and not on the angular momentum they carried.
Let the F string have total length $L_T=2\pi n_w R'$. Let it carry
$n_p$ units of momentum which implies the energy of vibration
\be
E={n_p\over R'}
\label{uone}
\ee
Further, let the wavelength of the vibrations be $\lambda \sim
L_T/n$. The Hamiltonian for small vibrations is
\be
H={T\over 2} \int _{z=0}^{z=L_T} dz [({d \vec x'\over d t})^2+({d\vec
x'\over d z})^2]
\ee
By the virial theorem
\be
\langle H\rangle=T \langle ({d\vec x'\over d z})^2\rangle L_T \sim T
\langle x'^2\rangle {2\pi \over n_w R'} n^2
\ee
Setting $\langle H\rangle = E$ we get
\be
\langle x'^2\rangle ^{1/2}\sim  ({N\over 2\pi T})^{1/2} {1\over n}
\ee
Thus we see that (for fixed $N$) the rough size of the singularity
depends on $n$, which determines the wavelength in the FP
solution, and maps to  the winding number of individual component strings in
the dual D1-D5 system. In the supergravity solution the size of
the singularity determines the approximate end of the throat, and we
see from the above analysis that for the D1-D5 system the
effective length of the throat is determined by the number of
component strings rather
than the angular momentum of the configuration.

For large $N$, the generic ground state of the FP system has $\sim
\sqrt{N}$ quanta of vibration modes with wavelength $\sim
L_T/\sqrt{N}$.  In the dual D1-D5 system we find that for the
geometries (\ref{MaldToCompare}) we have $\gamma\sim 1/\sqrt{N}$, which is zero
in the classical limit.

In Appendix \ref{AppenLen}  we compute the length of the singular curve for the
D1-D5 bound state in flat space. Multiplying this length with the
height $2\pi R$  of
the singularity in the $y$ direction we find for the area of this
classical singularity
\be
Area  \le  4\pi\sqrt{2}\sqrt{NG^{(6)}}
\ee
where equality is attained only for the cases where all component
strings have the
same length.

\subsection{Nonsymmetric excitations of component strings}

Let the CFT state have $m$ component strings. Note that the
calculation of Appendix A
for the amplitude of absorption/emission from the component string 
did not depend
on the length of the component string.    Thus an incident graviton
will create an excited state
of the form
\be
|\psi\rangle_{excited}={1\over \sqrt{m}}[(string ~1~ excited)+(string
~2~ excited) +\dots
+(string~m~ excited)]
\label{aone}
\ee

If all the $m$ component strings were   identical, then
(\ref{aone})  is the only state allowed
by Bose symmetry; this is the state that we used in the computation
of Appendix A.
But in general the $m$ component strings will not be in the same
state --  they may have
different lengths or different spin orientations. Then instead of
(\ref{aone}) we can get the
more general state
\be
|\psi\rangle_{excited}=\alpha_1(string ~1~
excited)+\alpha_2(string ~2~ excited) +\dots
+\alpha_m(string~m~ excited)
\label{atwo}
\ee
As a simple example, take the R ground state
$[\sigma_{N/m}^{--}]^{m/2}  [\sigma_{N/m}^{++}]^{m/2}$
and consider it excited to the nonsymmetrical combination
\be
|\psi\rangle_{excited}= {1\over \sqrt{2}}[(\sigma^{--}~string ~ excited)
        -(\sigma^{++}~string ~ excited)]
\label{athree}
\ee
A naive calculation would now suggest that the amplitude to emit a
graviton from this state
is zero -- the amplitude for emission from the two parts of the  state give
a total amplitude $  {1\over \sqrt{2}}[R-R]=0$ ($R$ is the amplitude
of emission from either
type of component string).   It would thus seem that there are CFT
excitations that
are decoupled from the supergravity modes at infinity.

But a closer look reveals that this naive computation requires the
following modification.
We have seen above that the singularity of the D1-D5 system is not a
point in the space $\vec x$ but
rather an extended curve. Even in the limit of weak coupling when we
have the `branes
in flat space', the location of the branes will be along this curve
of nonzero extent
(see Appendix \ref{AppenLen} for a computation of the length of the singular
curve). In
the emission calculation if
the component strings were all at one point then the graviton
emission amplitude would indeed
vanish for the state (\ref{athree}). But the finite extent of the
singularity implies that
even if the leading order s-wave emission is cancelled by a judicious
choice of phase as in (\ref{athree}), the excitation will decay
(though more slowly)
through the emission of p wave and higher partial wave gravitons.

We can see the presence of states analogous to (\ref{atwo}) in the
dual supergravity. A simpler example can be found using an unbound state rather
than a single bound state.  Thus consider two identical D1-D5 bound
states, each corresponding
to the CFT state $\sigma_N^{--}$. Construct a superposed state where
the centers of
the two components are placed a distance $b$ apart, with
$b\ll (Q_1Q_5)^{1/4}$. Then
the supergravity solution exhibits a single `combined' throat for a
certain distance,
and then the throat branches into   two smaller throats, which we
call throat 1 and throat 2.\footnote{Branching throats were
considered in a slightly different
context in \cite{barnchThr}.}

An s-wave scalar traveling down the combined throat enters throat 1
and throat 2 with equal amplitude.
But we can also construct a wavefunction for the scalar which has
{\it opposite} amplitudes
in throat 1 and throat 2. How will such a wavefunction emerge into
the  combined throat and thus
out to infinity?

While the above wavefunction does not emerge into the combined throat
as an s-wave,
it does have a nonzero amplitude to emerge as a p-wave. The p-wave
has a smaller amplitude to exit the combined throat,
and thus we see a suppression in the radiation rate for such a wavefunction.

In a general state of the D1-D5 system the end of the throat exhibits
no particular symmetry,
and so an s-wave going down the throat can get converted to a mixture
of harmonics after reflection
from the end of the throat. The geometries (\ref{MaldToCompare}) are special
however, since the different harmonics
separate exactly and an s-wave returns as an s-wave.  The
corresponding CFT states
are special too,
since their excitation takes the symmetrical form (\ref{aone}) and
not the more general form
(\ref{atwo}).

\section{Breakdown of the semiclassical approximation and the
threshold of formation
            of black holes}
\label{SectBreak}
\renewcommand{\theequation}{5.\arabic{equation}}
\setcounter{equation}{0}

\subsection{Several vibration modes on the same component string}

Consider an s-wave minimally coupled scalar, with energy $\omega$,
traveling up and down the
throat of the D1-D5 geometry. We have seen that this maps in the CFT
to a left moving and a
right moving vibration mode on the effective string, each with energy
$\omega/2$. In particular
we had noted (eqn.  (\ref{seven})) the maps
\bea
h_{67}&\rightarrow& {1\over \sqrt {2}}~(|x_6\rangle_L\times
|x_7\rangle_R+|x_7\rangle_L\times  |x_6\rangle_R)\nonumber \\
B^{RR}_{67}&\rightarrow& {1\over \sqrt {2}}~(|x_6\rangle_L\times
|x_7\rangle_R-|x_7\rangle_L\times  |x_6\rangle_R)
\eea
We will find it convenient to work with the linear combinations
            \bea
S^+_{67}&\equiv& {1\over \sqrt{2}}[h_{67}+B^{RR}_{67}]\rightarrow
|x_6\rangle_L\times
|x_7\rangle_R\nonumber \\
S^-_{67}&\equiv& {1\over \sqrt{2}}[h_{67}-B^{RR}_{67}]\rightarrow
|x_7\rangle_L\times
|x_6\rangle_R\nonumber
\eea

\b

{\subsubsection{Throat corresponding to the CFT state with one component
string.}}

            Consider the R ground state that
arises from the chiral primary
$\sigma_N^{--}$. This state has one `component string' wound $n_1n_5$
times around the circle
of radius $R$ (figure \ref{figGround}a), and thus has the longest possible
throat ($\Delta
t_{CFT}=\Delta t_{SUGRA}=\pi
n_1n_5 R$). Place in this throat a quantum of $S^+_{67}$ with energy
\be
\omega={2\over n_1n_5 R}\equiv \omega_0
\ee
This gives left and right vibrations on the effective string of
energy $1/n_1n_5 R$ each, which
is the lowest energy excitation possible.

Now imagine that we {\it also} place in the throat a quantum of
$S^+_{89}$ with energy
$\omega_0$. In the CFT we get the state
\be
S^+_{67}(\omega_0)S^+_{89}(\omega_0)\rightarrow |x_6\rangle_L
|x_8\rangle_L\times |x_7\rangle_R |x_9\rangle_R
\label{eone}
\ee
But now look at the CFT state corresponding to the supergravity
scalars $S^+_{69}(\omega_0),
S^+_{78}(\omega_0)$:
\be
S^+_{69}(\omega_0)S^+_{87}(\omega_0)\rightarrow |x_6\rangle_L|x_8\rangle_L
\times |x_7\rangle_R|x_9\rangle_R
\label{ytwo}
\ee
The supergravity states  in (\ref{eone}), (\ref{ytwo}) are different
(the scalars involved are not
the same in the two cases)
but they seem to map to the same state in the CFT. It would appear
that we have found a
contradiction with   the  proposed duality map.

But note that these quanta in the throat have  an energy of order
$\sim 1/n_1n_5 R$ or greater. It was   shown in \cite{lm4} that for
this throat geometry the
threshold of horizon formation is
\be
\Delta E_{threshold}\sim {1\over n_1n_5 R}
\ee
{\it Thus just when we seemed to be getting a contradiction between
the CFT effective string and the throat
geometry, we find that the physics changes because of black hole formation.}

           Note that
for this particular state of the CFT the supergravity approximation
will not be good even to
describe the propagation of one quantum to the end of the throat,
since the backreaction on the
geometry will deform the geometry by order unity.

\b

{\subsubsection{Throats corresponding to CFT states with $m$ component
strings.}}

Let us see how the above discussion extends to states of the CFT that
have $m>1$ component
strings. Let the CFT state be composed of $m$ component strings with
winding number $\sim N/m$ each. Then in the supergravity solution we see that
(eq. (\ref{ttwo})
\be
a\sim {m\over n_1n_5} \sqrt{Q_1Q_5} {1\over R}
\label{rone}
\ee
The throat ends at $r\sim a$ while for $r\gg a$ it has the geometry 
(\ref{qfive}) of
the D1-D5 system with no
rotation. In the  geometry  (\ref{qfive}) if we add a nonextremal energy
$\Delta E$ then we get a horizon
at
\be
r_H=\left[{2g^2\Delta E\over RV}\right]^{1\over 2}
\label{rtwo}
\ee
(we have set $\alpha'=1$).
Thus $r_H\gg a$ then we get horizon formation, while if $r_H\ll a$ then the
matter quanta just move
back and forth in the throat, giving the `hot tube' \cite{lm4}. The
threshold energy for black
hole formation $\Delta E_{threshold}$ is then obtained by setting
$r_H\sim a$, which  we write in
a suggestive fashion as
\be\label{ThreshEqn}
\Delta E_{threshold}\sim {m^2\over n_1n_5 R} = m ({1\over { N\over  m}R})
\ee
{\it We see that a black
hole forms if we have enough energy to excite one quantum  of the
lowest allowed vibration
mode on {\it each} component string.}

The minimum wavelength that fits in the throat is $(m\omega_0)^{-1}$. Note that
we can place $1<<m_1<<m$ quanta of this wavelength in the throat
without
  any significant deformation of the
throat geometry.  Thus
supergravity analysis performed in subsection 3.3 using the test
particle approximation is seen to be
valid, as long as we do not take too high an energy ($E>m^2\omega_0$) 
for this test
particle.

To summarize, we have seen that if we place more than one pair of
vibration modes on each component string,
then the naive count of supergravity states fails to agree with the
count in the CFT (eqs. (\ref{eone}, \ref{ytwo})). But the energy required to
excite the lowest vibration on each component string turns out to
equal $\Delta E_{threshold}$, the energy to form
a black hole; thus the physics changes at this point, and a
contradiction is averted.\footnote{In \cite{dmfolds} it
was found that in the $c=1$
matrix model ($D=2$ string theory) the process of `fold formation'
distributed an initial
energy $m^2$ as $\sim 1+2+\dots + m$; i.e. we get one
quantum each of frequencies
$1,2, \dots m$. It may be that the interactions in the
present problem distribute
energies
in this fashion rather than $m^2\sim m+m+\dots m$ as was suggested by
(\ref{asix}).}

\b
\subsection{Black hole formation}

Thus far we have ignored all interactions when dealing with the
`component strings' in the CFT; we just considered
left and right moving vibration modes on individual component
strings. But the supergravity solution does not in
general map to the CFT at the orbifold point $M^N/S_N$; we have a
deformation of the orbifold by blowup modes \cite{dj, dvv}.
Such deformations can be generated by $(1,1)$ operators of the form
$\psi\bar\psi \sigma_2$: the chiral primary
$\sigma_2$ has dimensions $({1\over 2}, {1\over 2})$ and $\psi,
\bar\psi$ are dimension ${1\over 2}$ fermions
from the left and right sectors. This operator cuts or joins
component strings, at the same time creating a left and a
right moving fermionic excitation on the resulting component strings
\cite{mang}.

To see what this interaction might do, let us look at the location of
the horizon $r_H$ (eqn. (\ref{rtwo}))  when we
throw in an energy $\Delta E\gg \Delta E_{threshold}$. (We still
assume that $\Delta E$ is low enough that the
horizon forms in the throat rather than outside.)
           From our map between the
supergravity throat and the
CFT we find that when a quantum travels from the start of the throat
to $r=r_H$ then on the
CFT string the two vibration quanta  separate by a distance equal to
the time of flight to $r_H$:
\be
l_H=(Q_1Q_5)^{1/2}\int^{ (Q_1Q_5)^{1/4}}_{r_H} {dr\over r^2} \sim
{(Q_1Q_5)^{1/2}\over r_H}
\ee
This requires the CFT string to have length at least $2l_H$.  When a
black hole forms we must clearly expect some
change in the description on the CFT side. Using (\ref{rtwo}) we find that
\be
\Delta E \sim  {4\pi\over 2l_H} ({2\pi n_1n_5 R\over 2l_H})
\label{asix}
\ee
{\it Thus we see that if the interaction were to change the CFT state
to one where all the component strings had
length
$2l_H$, then the energy $\Delta E$ would excite the lowest allowed
vibration energy ${4\pi\over 2 l_H}$ on each of the resulting ${2\pi n_1n_5
R\over 2l_H}$ component strings.}

\b

           Even though we are not investigating the interaction in detail here,
there are two features of the interaction that
deserve comment. First, below the black hole formation threshold we
have obtained good results by ignoring the
interaction altogether.\footnote{This may no longer be true when we have a
bound state with component strings of different lengths. The shape of
the singularity is obtained by mapping to the dual FP system where the F string
carries harmonics of two different orders. The resulting singularity
shape depends
in a complicated way on all the harmonics present, so we expect some
interaction effects between the component strings in the CFT.}
Secondly, when a
horizon does form (and thus interactions presumably become relevant) then we
seem to need enough energy to excite {\it all} the component strings.
We speculate
now on how this `all or none' feature of the interactions might arise.

In the CFT the normalized interaction operator has the form
\be
\psi\bar\psi\sigma_2\sim {1\over
N}\sum_{i,j=1}^N\psi\bar\psi\sigma_{ij}
\label{jone}
\ee
Let us start with a CFT state which has $n_1n_5$ singly wound
component strings, and let the first string carry a
pair of  vibrations. The above interaction can join another string to
the first one, giving a state of the form
\be
|\psi_f\rangle \sim {1\over \sqrt{N-1}}[(12)+(13)+ \dots (1N)]
\label{jtwo}
\ee
 From (\ref{jone}), (\ref{jtwo})  we see that the interaction generates a change
\be
|\psi_f\rangle \sim {1\over \sqrt{N}} |\psi_i\rangle
\ee
This small parameter ${1\over \sqrt{N}}$ is also the coupling
constant of supergravity in $AdS_3$, so it appears
that ignoring interactions in the CFT  was equivalent to ignoring gravity
corrections at leading order in the dual supergravity.

If we use
several such interactions, then we get further suppression by powers
of  ${1\over \sqrt{N}}$. But now suppose that
we convert all the initial component strings to component strings of
length 2, each carrying, say, the lowest
vibration mode. Then because all these component strings will be in
the same state, we get a `Bose enhancement
factor' -- if there are already $\sim N$ units of a given component
string then creating the next one gives a factor
$\sim \sqrt{N}$. Thus while the amplitude may be small to create a
few interactions, it may be order unity if it
affects almost all the component strings, which is what we observed
in the process of horizon formation.\footnote{The issue of a `phase transition' at the 
threshold of black hole
formation was discussed in \cite{m5}.}

\subsection{Validity of the supergravity approximation}

We have noted after eqn. (\ref{ThreshEqn}) that if the number of component
strings is $m\gg 1$ then the backreaction of a quantum with
energy $E\ll m^2\omega_0$ will not make a significant distortion to
the geometry ($m^2\omega_0=\Delta E_{threshold}$, the
threshold for black hole formation). The geometry of the throat is
(except near the end) locally $AdS_3\times S^3\times M$; since
this is an exact string solution it is not modified by stringy
corrections. As we go down the throat the sizes of $S^3$ and $M$ do not
change, but the length of the $y$ circle shrinks. One might wonder if
this leads to a large effective coupling constant for quanta
deep in the throat, thus possibly invalidating the linear wave
equation we have used for the propagation of the scalar. But in
\cite{lm4} it was shown that if the scalar has for instance an
interaction $\phi^3$ then this interaction becomes relevant only after
a  distance down the throat corresponding to a time of flight $ \sim
n_1n_5 R$. But the end of the throat
is reached after the quantum travels for a time $\sim n_1n_5 R/m$.
Thus again we find that if
$m\gg 1$ the interactions can be ignored, and we can use the linear
wave equation to explore the end of the throat at $r\sim a$.

\section{A proposal to resolve the information paradox}
\renewcommand{\theequation}{6.\arabic{equation}}
\setcounter{equation}{0}

In \cite{mes} it was argued that the information paradox is resolved
if we assume that spatial
slices in a foliation cannot be {\it stretched} too much. We review
this argument, and then relate
it to the computations of the present paper.

\b

{\subsection{The argument.}}

The information paradox arises because we can find a smooth region in
the  black hole spacetime
which  can be given a `nice' foliation by spacelike hypersurfaces;
         the spatial slices satisfy all the smoothness conditions that
we might wish to impose. On
the initial slice we have a regular distribution of matter, while
late time slices  capture both the
initial matter and the corresponding outgoing Hawking radiation. (The
slices do not approach
the singularity.) Since information cannot be `duplicated', we must
somehow have `bleached' the
information out of the matter and transferred it to the radiation, in
the course of the evolution.
But before we can postulate the existence of some  nonlocal process
to accomplish this goal we
must identify the `trigger' for the process: what distinguishes the
smooth foliation of this region of the black
hole spacetime from the foliation of a smooth manifold without black
holes? In the absence of
such a `trigger' criterion, the postulated nonlocal process would
invalidate normal quantum
evolution even in the absence of black holes.

Since the foliation is smooth, the trigger cannot involve short
distance quantum gravity effects.
Two features are however observed in the smooth black hole
foliations.  The first is that  one
part of the spacelike slice evolves much slower than the other. This
is the `redshift effect' familiar
in black hole geometries. But if we consider a scalar field evolving
along the foliation, we see no
reason why this differential evolution should invalidate normal
quantum evolution.  String
theory is more complicated, but here again there is no convincing
reason why the differential
evolution rates would lead to nonlocal information transport
(see however \cite{sosspol}). These
are the standard difficulties
that one runs into when trying to resolve the information problem.

The second feature of the slices  is that  they {\it stretch}
exponentially in the course of the evolution. More precisely, if we
want the slices to be smooth
(thus they should not go near the singularity) then the price that we
must pay is that a
region of the initial slice (with radius $\sim R_s$, the Schwarzschild radius)
will get `stretched' to a final
length that is $\sim O(1/\hbar)$ and thus nonclassical.

\begin{figure}
\epsfxsize=5.5in \epsffile{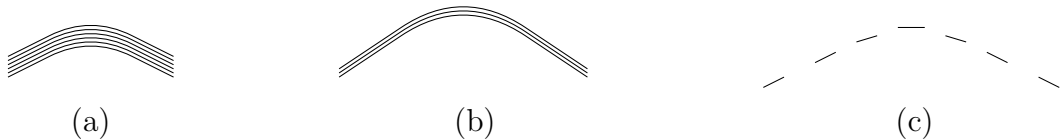}
\begin{picture}(400,20)
\put(23,5){(a)}
\put(168,5){(b)}
\put(335,5){(c)}
\end{picture}
\caption{(a) An initial hypersurface with a high density of degrees
of freedom; ~ (b) Stretching dilutes
these degrees; ~ (c) The degrees of freedom are so sparse that
usual local physics breaks down.\label{FigStrLin}}
\end{figure}

It was argued in \cite{mes} that spatial slices should be attributed
a `density of degrees of
freedom'. Stretching dilutes these degrees. If we place more matter
quanta in a region of the
slice than there are available degrees of freedom then semiclassical
evolution breaks down and
nonlocal information transport can be triggered (Figure \ref{FigStrLin}).
This new way of
violating the semiclassical
approximation allows information leakage in Hawking radiation while
preserving locality of
quantum mechanics in the absence of black holes, and thus the
information paradox would be
resolved. The length of this critically stretched slice was expressed
in \cite{mes} in terms of the number of degrees of freedom
$e^S$ ($S$ is the black hole entropy), but noting that the separation
of these degrees along the slice was order the Schwarzschild
radius $R_s$, we can write the critically stretched length of the slice as
\be\label{StrSmax}
      s_{max}\sim {{\cal V}\over G_N}
\label{ione}
\ee
where ${\cal V}$ is the volume enclosed in {\it flat} space in a ball
of radius $\sim R_s$.

The above arguments can be made on very general grounds, reflecting
the universal nature of
the information problem. Let us now see what aspects of this picture
emerge in our D1-D5
system.

\b

{\subsection{`Stretching' of slices in the D1-D5 system}}

\begin{figure}
\begin{center}
\epsfysize=2in \epsffile{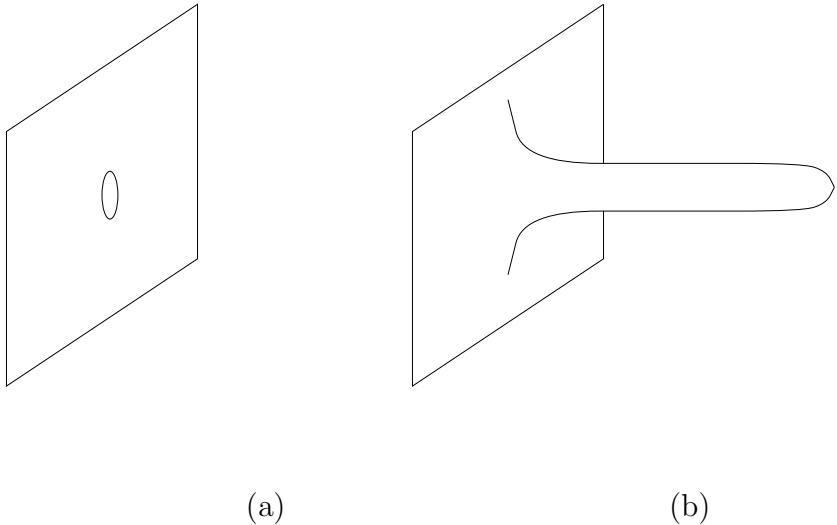}
\end{center}
\begin{picture}(350,20)
\put(90,0){(a)}
\put(250,0){(b)}
\end{picture}
\caption{\label{figStretch}(a) Flat spacetime in the absence of the
branes; ~ (b)
The presence of branes stretches the region marked in (a) to a long throat.}
\end{figure}
(i) \q In the absence of the D1, D5 branes the spacetime geometry is
flat (figure \ref{figStretch}a). The branes
deform the geometry to a long but finite throat. Let us characterize
the `stretching' of the
geometry by the time delay $\Delta t_{SUGRA}$ caused by propagation
down and back up   the
throat.  We have found that the maximum value of this time delay is
\be
\Delta t_{max}={\pi Rn_1n_5}
\label{wone}
\ee

We wish to relate this length scale (\ref{wone}) to the size of the
region {\it before} stretching.
The radius $\hat r$ of the ball drawn in figure \ref{figStretch}b is $\sim
(Q_1Q_5)^{1/4}$. The radius of the $S^3$
stabilizes to a constant value in the throat (until we near the end),
and for concreteness we set
$\hat r$ to this radius
\be
\hat r=(Q_1Q_5)^{1/4}
\ee
For thus value of $\hat r$ the enclosed volume in {\it flat} space
(i.e. the space before
deformation) is
\be
{\cal V}={\pi^2\over 2}( \hat r)^4={\pi^2\over 2} Q_1Q_5
\ee
We now note that
\be\label{StrTmax}
\Delta t_{max}={{\cal V}\over 2G_N^{(5)}}
\label{wtwo}
\ee
where $G^{(5)}_N={G^{(10)}_N\over (2\pi)^5 VR}$ and
$G^{(10)}_N=8\pi^6 g^2 \alpha'^4$.  We observe that the relation 
(\ref{wtwo}) has the
same form as (\ref{ione}).

It is
noteworthy that the individual quantities in (\ref{wtwo}) depend on
$g, \alpha', V, R, n_1, n_5$
but these parameters appear only in certain physically intuitive
combinations in the final
relation. This fact leads one to believe  that the relation
(\ref{wtwo}) has a validity beyond the
specific D1-D5 system being studied here.

\b

(ii) \q Let the state in the D1-D5 CFT be given by $\sim m$ component
strings of winding number
$\sim N/m$ each. The time delay in the supergravity throat is then
$\Delta t_{SUGRA}\sim
\Delta t_{max}/m$. We have seen in section \ref{SectBreak} that if more than
$\sim m$ supergravity quanta
are placed in this throat geometry then we get evolution that departs
from naive semiclassical
expectations. {\it This suggests that a throat stretched to have a time
delay $\Delta t_{max}/m$ has
just enough degrees of freedom to support $\sim m$ quanta while maintaining the
supergravity approximation.} This indicates a `dilution' of the
degrees of freedom when the throat is `stretched'
to longer lengths.

\b

(iii)\q The argument of \cite{mes} requires that is we try to put
more data on a slice
than there are degrees of freedom then we will trigger nonlocal
effects. Let us examine a similar feature that we
find in our present study.  Let the supergravity throat contain the two quanta
$S^+_{67}, S^+_{89}$. Instead of
putting these  quanta in the lowest energy states
in the throat let them have reasonably localized positions $r_1\ne
r_2$. Let the CFT state have only a
few component strings.
Then from the form of the wavefunctions (\ref{aone}) we see that
there is a significant  amplitude to have both
pairs of vibrations on the same component string.   But
this CFT state can be regarded, a seen above in section
\ref{SectBreak},  as describing with some amplitude the
supergravity quantum  $S^+_{69}$. If we try to extract
the wavefunction for $S^+_{69}$ however, we find that this supergravity quantum
         is not localized near
either
$x_1$ or $x_2$.

\bigskip

To summarize, while we have not analyzed in the present paper the
same spatial slices as those considered in \cite{mes}, we have
seen analogous phenomena of `stretching' and consequent `dilution' of
degrees of freedom; further, the relations expressing the
maximum stretch of spacelike slices (\ref{StrSmax}), (\ref{StrTmax})
appear to be similar.

\b

\section{Discussion}
\renewcommand{\theequation}{7.\arabic{equation}}
\setcounter{equation}{0}

Let us summarize our results and note their relation  to earlier work.

In \cite{mal, gkp} the Poincare patch arising from the near horizon
geometry was considered. It was noted
that  geometry stopped being $AdS$    at large $r$ ($r>
(Q_1Q_5)^{1/4}$).  The naive  geometry eqn. (\ref{qfive})  for the
D1-D5 system (without rotation)  extends as a uniform throat all the
way down to a
horizon at $r=0$. What we have noted is that the throat  ends at
a certain point, and the location and nature of this end of the throat mirrors
the D1-D5 microstate of the dual CFT.  In the process we found
that the metrics of
\cite{mm} were only a small family out of a general family of metrics
for the D1-D5 system, and we studied these more general
metrics using chiral null models.

Having this end to the throat was crucial to all our considerations
-- if we took
instead the naive geometry (\ref{qfive}) then all quanta  falling down
the throat
would reach the horizon at $r=0$ and create a  black hole in the
process. The actual ending of the throat implies a minimum threshold
$\Delta E_{threshold}$ for black hole formation, and yielded an energy  domain
below this threshold where we could set up a detailed duality map
to the  CFT.  The finite length of the throat is essential to
obtaining the
relation $\Delta t_{CFT}=\Delta t_{SUGRA}$. In the analysis of 
\cite{hm} a quantum
falling down the throat of the 3-brane geometry was represented by a spreading
wavepacket in the CFT, but there was no analogue of the relation 
(\ref{tfive}) since the
infalling quantum did not turn back without horizon formation.

The finite length of the throat arises both from the truncation at
small $r$ ($r\sim
a$) and at large $r$ ($r\sim (Q_1Q_5)^{1/4}$).  We used explicitly
the geometry at
$r\sim (Q_1Q_5)^{1/4}$  (where the throat joins flat spacetime) in
computing the
radiation rate ${\cal R}_{SUGRA}$ from the throat, and found that this
equaled the radiation rate ${\cal R}_{CFT}$ from the CFT state.  Note that our
computations were somewhat different from earlier studies \cite{DasTriv}
where the join of
the AdS region to flat space was made at an arbitrary location in the AdS;
     the operator breaking conformal symmetry in the CFT then depended on this
location.

It is remarkable that the black hole threshold for different
microstates depends only on the energy scale $1/R$  and the winding
numbers of the component strings in the CFT. We noted that a
breakdown of the semiclassical approximation occurred when there
was more than one pair of vibration modes on the same component
string. Such a proposal would  appear to be closely related to
the stringy exclusion principle, but there are some important differences
which we now discuss.

The microscopic D1-D5 CFT is manifestly unitary and the degrees of
freedom are limited. The information paradox
asks for the implication of
this fact in supergravity. It has often been argued
that the `stringy exclusion principle' \cite{exclusion}
implies that there is a cut off in the order of spherical harmonics
on the $S^3$. While this is
probably true, it has  not yielded any direct implications for the
information problem, in part
because the Hawking radiation is mostly confined to the first few
spherical harmonics. What we
have suggested here is that the cutoff of degrees of freedom in the
CFT implies a maximal radial
stretching of slices in the geometry, and the arguments of \cite{mes}
then relate this to a
resolution of the information paradox.

A further essential difference is that we are looking at non-BPS
excitations, rather
than the   BPS quantities studied in \cite{exclusion, db}. If we
consider a  component string wound $n$ times on a circle of
radius $R$, then left and right vibrations occur in fractional units
but the total units of momentum must be an integer. Thus the
lowest non-BPS excitation has energy ${2\over nR}$,  while the lowest
BPS excitation has (for large $n$)  a much higher energy
${1\over R}$.  Correspondingly, we find have found very low black hole
thresholds for CFT states which have just one long component string.
By contrast the black hole threshold discussed in \cite{db} pertains
to the NS sector of the CFT,  and is a single,  high energy
threshold  rather than a collection of thresholds depending on the
particular CFT microstate.

Perhaps most interesting is the fact that we have  found some support
for the proposal
of
\cite{mes} that if we `stretch' a spatial
hypersurface too much then we will encounter a breakdown of normal
semiclassical evolution, even though we would
not see any of the usual factors that violate the semiclassical
approximation like planckian curvature or planckian energies.
It is interesting that the order of magnitude relation (\ref{ione}) 
emerging from
\cite{mes} has the same form as the more precise relation
(\ref{StrTmax}), though we must note that the hypersurfaces involved in
(\ref{StrSmax}) and (\ref{StrTmax}) are somewhat different. If
the relation (\ref{StrTmax}) turns out to be a general result, as we have
conjectured, then we would find a profound change in our
understanding of the ultimate structure of spacetime.

\section*{Acknowledgments}

We are grateful to T.~Banks, I.~Klebanov, F.~Larsen and A.~Strominger
for useful
discussions. This work was supported in part by  DOE grant DE-FG02-91ER40690.

\appendix

\section{Rate of radiation from the effective string}
\renewcommand{\theequation}{A.\arabic{equation}}
\setcounter{equation}{0}

Consider a string wrapped around the direction $x_5\equiv y$. There is one
right moving mode and one
left moving mode on the string, with energy-momentum vectors given by
(\ref{six}).
            The polarizations of the vibration modes are given by the
wavefunction (\ref{seven}).
We wish
to find the probability per unit time for these modes to interact and
emerge from the string as a
graviton.

The vertex that couples a left vibration mode, a right vibration
mode, and a graviton is found
by using the DBI action for the string \cite{dm1, dm2}. We first place the
   entire
system in a large box. The
volume of the noncompact directions $x_1, x_2, x_3, x_4$ is $V_{nc}$,
the volume of the compact
$T^4$ (in the directions $x_6, x_7, x_8, x_9$) is $(2\pi)^4 V$, and
the length of the $x_5$ circle
is $2\pi R$. Thus the volume of the 9-dimensional space is
$V_9=(2\pi)^4 V (2\pi) R
V_{nc}$.  We set $\alpha'=1$. The 10-d Newton's constant is
$G_N=8\pi^6 g^2$, and $8\pi G_N=\kappa^2$.

First consider the case where we have just one component string
wrapped $n_1n_5$ times around $y$.
For such a  string in \cite{dm2} we computed the amplitude per unit
time for the state
$|6\rangle_L\times
|7\rangle_R$ to decay into any given Fourier mode of the graviton $h_{67}$.
This quantity is
\be
\tilde R_h=\sqrt{2}\kappa |p_5|{1\over \sqrt{2\omega}}{1\over \sqrt{V_9}}
\ee
where $p_5$ is the momentum of any one of the vibration modes along
$x_5$, and $\omega$ is
the energy of the graviton. For the state (\ref{six}) of the
vibration modes the amplitude per
unit time will be higher by a factor $\sqrt{2}$. Using that
$|p_5|=\omega/2$ we get for the
amplitude of decay per unit time
\be
R_h=2\kappa |p_5|{1\over \sqrt{2\omega}}{1\over \sqrt{V_9}}
\label{oone}
\ee
Using the `Fermi golden rule'  we find the probability of emission
per unit time
\be
{\cal R}={2\pi |R_h|^2 \over \Delta E}
\ee
where $\Delta E$ is the spacing between levels for the graviton. The
graviton has momentum
only in the noncompact directions $x_1, \dots x_4$. Thus
\be
\Delta E= {(2\pi)^4\over V_{nc} }{1\over 2\pi^2 \omega^3}
\ee

Putting all this together we finally get for the probability of
emission per unit time
\be
{\cal R}={\pi^2\omega^4 g^2\over 2 V (2\pi) R}
\ee

Now consider the R ground state arising from the chiral primary
$[\sigma_{N/m}^{--}]^m$. There are $m$ component strings, each in the
same state.
Because of Bose symmetry between identical component strings, the
wavefunction for
       the excited system must have the form
\be
|\psi\rangle = {1\over \sqrt{m}}[(string ~ 1~ excited) + (string ~ 2
~ excited) +\dots (string ~ m ~ excited)]
\ee
The amplitude $R_h$ is now
\be
R_h=\sqrt{m}2\kappa |p_5|{1\over \sqrt{2\omega}}{1\over \sqrt{V_9}}
\ee
(there are $m$ contributions each weighted by ${1\over \sqrt{m}}$).
The probability per unit time for decay is
then
\be
{\cal R}=m{\pi^2\omega^4 g^2\over 2 V (2\pi) R}
\ee

\section{Rotating D1--D5 system.}
\label{AppD1D5Rev}
\renewcommand{\theequation}{B.\arabic{equation}}
\setcounter{equation}{0}

In this appendix we review some properties of the geometries
(\ref{MaldToCompare}). These metrics preserve a $U(1)\times U(1)$ subgroup out
of the $SO(4)$ rotational symmetry of the $S^3$.  As we have seen
in this paper, these metrics  represent only a small subclass of
all the ground states of the D1-D5 system, but they are useful
since the scalar wave equation factorizes in these backgrounds.

In supergravity approximation the system is parameterized by two
charges $Q_1$, $Q_5$ and the rotation parameter $a$. The 10-D
metric in the string frame is given by\footnote{In \cite{lm3} we
wrote the metric for rotating F1--NS5 system, but (\ref{D1D5}) can
be obtained from it by applying the S duality transformation.}:
\bea\label{D1D5}
ds^2&=&-\frac{{f_0}}{\sqrt{f_1f_5}}(dt^2-d{
y}^2)+ \sqrt{f_1f_5}(\frac{dr^2}{r^2+{ a}^2}+d\theta^2)+
\sqrt{\frac{{f_1}}{{f_5}}}\sum_{i=1}^4 dz^i dz^i\nonumber\\
&+&\frac{\sqrt{f_1f_5}}{f_0}\left[(r^2+\frac{{ a}^2Q_1Q_5\cos^2\theta}
{f_1f_5})\cos^2\theta d\psi^2
+(r^2+{ a}^2-\frac{{ a}^2Q_1Q_5\sin^2\theta}{f_1f_5})
\sin^2\theta d\phi^2\right]\nonumber\\
&-&\frac{2Q_1Q_5{a}}{\sqrt{f_1f_5}}\left[\sin^2\theta dtd\phi+
\cos^2\theta d{ y}d\psi\right]
\eea
This system also has a nonzero
value of dilaton field and the RR two--form $C^{(2)}_{\nu\nu}$:
\bea
e^{2\Phi}&=&\frac{{ f}_1}{{ f}_{5}},\qquad
C^{(2)}_{ty}=-\frac{2Q_1}{{ f}_{1}}, \qquad
C^{(2)}_{t\psi}=-\frac{\sqrt{Q_1Q_5}{ a}\cos^2\theta} {{ f}_{1}},
\nonumber\\
\label{FNSend}
C^{(2)}_{y\phi}&=&-\frac{\sqrt{Q_1Q_5}{ a}
\sin^2\theta}{{ f}_{1}},
\qquad
C^{(2)}_{\phi\psi}=Q_5\cos^2\theta+
\frac{Q_5{ a}^2\sin^2\theta\cos^2\theta}
{{ f}_{1}}.
\eea
Here we have introduced three convenient functions:
\be\label{f0Def}
{ f}_0=r^2+{ a}^2\cos^2\theta,\qquad f_1=f_0+Q_1,\qquad f_5=f_0+Q_5.
\ee
The coordinates $z^1,\dots,z^4$ form a torus with volume $(2\pi)^4V$, and
$y$ is compactified on a circle with circumference $2\pi R$.

>From the microscopic point of view the system (\ref{D1D5}) describes a
configuration of $n_1$ D1 and $n_5$ D5 branes with angular momentum $J$:
\bea\label{DefCharge}
n_1&=&\frac{V}{4\pi^2{g}}\int_{S^3}
e^{-\Phi}*H =\frac{Q_1 V}{g},
\nonumber\\
n_5&=&\frac{1}{4\pi^2}\int_{S^3} H =\frac{Q_5}{g},\nonumber\\
j&=&\frac{aVR}{2g^2}\sqrt{Q_1Q_5}
\label{mafour}
\eea
Here $g$ is the string coupling constant and we always put $\alpha'=1$.

In appendix \ref{AppGener} we will also need the relation between the solution
(\ref{D1D5}) and the solution for the rotating fundamental string carrying
momentum.
This relation
was studied in detail in \cite{lm3}, here we just mention some facts which
will later be used. Using the chain of string dualities one can map
(\ref{D1D5})
into the solution describing the fundamental string which is wrapped $n_5$
times around the $y$ circle, it also carries $n_1$ units of momentum and $J$
units of angular momentum. The metric and matter fields of the solution
describing the fundamental string can be written in terms of the chiral null
model. To do this one should first go to the coordinates ${\tilde r},
{\tilde\theta}$
instead of $r,\theta$:
\be\label{coord1}
{\tilde r}=\sqrt{r^2+a^2\sin^2\theta},\qquad
\cos{\tilde\theta}=\frac{r\cos\theta}{\sqrt{r^2+a^2\sin^2\theta}},
\label{maone}
\ee
and then introduce the Cartesian coordinates in the ${\tilde r},
{\tilde\theta},\phi,\psi$
space:
\bea\label{FlatCoord}
&&x_1={\tilde r}\sin{\tilde\theta}\cos\phi,\qquad
x_2={\tilde r}\sin{\tilde\theta}\sin\phi,\nonumber\\
&&x_3={\tilde r}\cos{\tilde\theta}\cos\psi,\qquad x_4={\tilde r}\cos
{\tilde\theta}\sin\psi,
\eea
as well as two null coordinates:
\be\label{NullCoord}
u=t+y,\qquad v=t-y.
\ee
In the coordinates (\ref{FlatCoord}), (\ref{NullCoord}) the geometry of the
rotating string reads (see \cite{lm3} for details):
\bea\label{FPnullCoor}
ds^2&=&e^{2\Phi}\left(-du'~dv'+\frac{Q'_1}{Q'_5}(e^{-2\Phi}-1)d{v'}^2-
4a'\sqrt{\frac{Q'_1}{Q'_5}}(e^{-2\Phi}-1)\frac{x'_1dx'_2dv'-x'_2dx'_1dv'}
{{\vec x'}\cdot{\vec x'}+{a'}^2+f'_0}\right)
\nonumber\\
&+&d{\vec x'}d{\vec x'}+d{\vec z'}d{\vec z'},\\
B_{uv}&=&G_{uv},\quad
B_{vi}=-G_{vi},\quad
e^{-2\Phi}=1+\frac{Q'_5}{f'_0}. \nonumber
\eea
Here rescaled coordinates and charges are defined by:
\bea
x_i'&=&x_i\frac{R\sqrt{V}}{g},\qquad u'=u\frac{R\sqrt{V}}{g},\qquad
z'_1=\frac{z_1\sqrt{V}}{gR_1},\qquad 
\\
Q'_1&=&\frac{Q_1R^2V}{g^2},\qquad Q'_5=\frac{Q_5R^2V}{g^2}, \qquad
a'=\frac{aR\sqrt{V}}{g},\\
\label{FpToD15R}
R'&=&\sqrt{V},\qquad V'=\frac{V}{gR_1^2},\qquad g'=\frac{VR}{gR_1},\qquad
R_1'=\frac{\sqrt{V}}{g}.
\eea
Here $f'_0=f_0\frac{R\sqrt{V}}{g}$ (not the derivative of $f_0$!),
           and in terms of the
new coordinates it is given by
\be
           f'_0=\left(({\vec x'}\cdot{\vec
x'})^2+
2{a'}^2({x'}_3^2+{x'}_4^2-{x'}_1^2-{x'}_2^2)+{a'}^4\right)^{1/2}.
\ee The geometry (\ref{FPnullCoor}) is an example of the chiral
null model, we will discuss some properties of such models as well
as their relations to the D1--D5 systems in the next appendix.

\section{A chiral null model approach to D1-D5 solutions}
\label{AppGener}
\renewcommand{\theequation}{C.\arabic{equation}}
\setcounter{equation}{0}

In  appendix \ref{AppD1D5Rev} we have seen that under a chain of
string dualities the rotating D1--D5 system (\ref{D1D5}) is mapped
to the rotating fundamental string carrying  momentum charge. The
resulting solution (\ref{FPnullCoor}) belongs to the class of the
chiral null models \cite{chnull}. Since chiral null models have
been used in the past as a powerful tool for generating new
solutions in  supergravity, we will devote this appendix to
studying such models and their transformations under the chain of
dualities relating the fundamental string (\ref{FPnullCoor}) and
the D1--D5 system (\ref{D1D5}).

Let us first recall some of the properties of chiral null models
\cite{chnull}.
The metric and matter fields for such models are given by:
\bea\label{ChiralNull}
ds^2&=&H'({\vec x'},v')\left(-du'~dv'+K'({\vec x'},v')d{v'}^2+
2A'_i({\vec x'},v')dx'_i dv'\right)+
d{\vec x'}\cdot d{\vec x'}+d{\vec z'}d{\vec z'},\nonumber\\
B_{uv}&=&-G_{uv}=\frac{1}{2}H'({\vec x'},v'),\qquad
B_{vi}=-G_{vi}=-H'({\vec x'},v')A'_i({\vec x'},v'),\\
&&\qquad\qquad e^{-2\Phi}=H'^{-1}({\vec x'},v').\nonumber
\label{ChiralSolution}
\eea
Regarding $A'_i$ as a gauge field we can construct the field strength
${\cal F}_{ij}=A'_{j,i}-A'_{i,j}$. The
functions in the chiral null model are required satisfy the equations
\be\label{NullEqn}
\partial^2 H'^{-1}=0,\qquad \partial^2 K'=0,\qquad \partial_i{\cal F}^{ij}=0.
\ee
Here
$\partial^2$ is  the Laplacian
in the $x'_i$ coordinates. Note that the indices $i,j$ span the
subspace $\{ x_i\}$ where the metric is just
$\delta_{ij}$, and thus these indices are raised and lowered by this
flat metric.

Note that the geometry (\ref{FPnullCoor}) has a form of the chiral null
model with
\bea\label{myChiral}
H'^{-1}=1+\frac{Q'_5}{f'_0},&&\qquad K'=\frac{Q'_1}{f'_0},\nonumber\\
A'_1=\frac{2\sqrt{Q'_1Q'_5}a'x'_2}{f'_0(f'_0+{\vec x'}{\vec x'}+a'^2)},&&
A'_2=-\frac{2\sqrt{Q'_1Q'_5}a'x'_1}{f'_0(f'_0+{\vec x'}{\vec x'}+a'^2)},~~
A'_3=A'_4=0 .
\eea

We now wish to develop an analogue of the chiral null models
directly for the D1-D5 system; i.e., we wish to write supergravity
solutions to the D1-D5 system in terms of functions that can be
linearly superposed to generate new solutions. We can obtain such
a formalism by applying dualities to the chiral null modes
describing the FP system. But at some stage in these dualities we
will have to perform a T-duality along the direction $y'$, so we
look at FP solutions that are independent of $y'$ to start with.
This means that the functions $H'$, $K'$ and $A'_i$ do not depend on
$v'$, but are only functions of ${\vec x}'$.

First we make an S duality transformation to make the
original string into the D1 brane carrying momentum. Then by applying T
dualities along all directions of the torus ($z_1,z_2,z_3,z_4$) we produce
the D5 brane carrying momentum:
\bea\label{D5Pmetr1}
ds^2&=&-H^{1/2}(dt^2-dy^2)+KH^{1/2}(dt-dy)^2+2H^{1/2}A_idx^i(dt-dy)\nonumber\\
&+&H^{-1/2}d{\vec x}d{\vec x}+H^{1/2}d{\vec z}d{\vec z}\phantom{\frac{a}{b}}
\\
\label{D5Pmetr2}
e^{2\Phi}&=&H,\\
\label{D5PR6}
C^{(6)}_{ti6789}&=&C^{(6)}_{iy6789}=-HA_i,\qquad
C^{(6)}_{ty6789}=H-1
\eea
Since we will not use this metric later on, we are not writing a
rescaling of coordinates which should be done to go from (\ref{ChiralNull}) to
(\ref{D5Pmetr1}).

At this stage it is convenient to describe the RR fields not in terms of the
six form (\ref{D5PR6}), but in terms of the dual two form $C^{(2)}$. To
construct this form we apply the electric--magnetic duality to (\ref{D5PR6}).
The field strength corresponding to (\ref{D5PR6}) has following nonzero
components:
\be
G^{(7)}_{tij6789}=-G^{(7)}_{ijy6789}=
\d_i(HA_j)-\d_j(HA_i),\qquad
G^{(7)}_{tiy6789}=-\d_i H
\ee
By applying the Hodge duality in ten dimensions we get a magnetically dual
field strength:
\bea\label{EMDuality}
G^{(3)}_{ijk}&=&-\eps^{ijkl}\d_l H^{-1},\\
G^{(3)}_{tij}&=&-G^{(3)}_{ijy}=\eps^{ijkl}\d_k A_l
\eea
This field strength corresponds to the two--form RR field $C^{(2)}_{\mu\nu}$:
\be
G^{(3)}_{\mu\nu\la}=3\d_{[\mu}C^{(2)}_{\nu\la]}
\ee
So far we were only looking at configuration with no dependence upon $t,y,z_i$,
so it is natural to require that $C^{(2)}_{\nu\la}$ has the same property.
This requirement is equivalent to fixing the gauge in $C^{(2)}$.

>From the structure of $G^{(3)}$ we conclude that the only nontrivial
components of $C^{(2)}$ are
\be
C^{(2)}_{ij}\equiv C_{ij},\qquad
C^{(2)}_{ti}=C^{(2)}_{iy}\equiv B_i
\ee
and in terms of these new fields equations (\ref{EMDuality}) become:
\be\label{DualFields}
dC=-^*dH^{-1},\qquad dB=-^*dA.
\ee
Here Hodge dual is taken with respect to the four dimensional space
$x^1,x^2,x^3,x^4$ with flat metric. Note that due to the equations of motion
for the null chiral model (\ref{NullEqn}):
\be
d^*dH^{-1}=0, \qquad d^*dA=0
\ee
the equations (\ref{DualFields}) can be integrated to give the forms $C$ and
$B$.

To summarize, the chiral null model dualized to the D5--P solution has metric
(\ref{D5Pmetr1}) and dilaton field (\ref{D5Pmetr2}), but the RR fields can be
described in
two alternative ways. We either have a RR 6--form (\ref{D5PR6}) or the
RR 2--form
\be\label{D5PR2}
C^{(2)}_{ij}= C_{ij},\qquad
C^{(2)}_{ti}=C^{(2)}_{iy}=B_i
\ee
which is related to (\ref{D5PR6}) by the electric magnetic duality
(\ref{EMDuality}).

We can now apply S duality followed by T dualities along $y$ and $z_1$
to transform the solution (\ref{D5Pmetr1}), (\ref{D5Pmetr2}),
(\ref{D5PR2}) into the
F1--NS5 system. Application of another S duality gives the D1--D5 solution:
\bea\label{D1D5Chiral}
ds^2&=&\sqrt{\frac{H}{1+K}}\left[-(dt-A_idx^i)^2+(dy+B_idx^i)^2\right]
+
\sqrt{\frac{1+K}{H}}d{\vec x}\cdot d{\vec x}\nonumber\\
&+&\sqrt{H(1+K)}d{\vec z}\cdot d{\vec z}\\
e^{2\Phi}&=&H(1+K),\qquad
C^{(2)}_{ti}=\frac{B_i}{1+K},\qquad
C^{(2)}_{ty}=-\frac{K}{1+K},\nonumber\\
C^{(2)}_{iy}&=&-\frac{A_i}{1+K},\qquad
C^{(2)}_{ij}=C_{ij}
\eea
The functions $H$, $K$ and $A_i$ appearing in this solution have the same
values as $H'$, $K'$ and $A'_i$:
\be
H({\vec x})=H'({\vec x'}), \qquad K({\vec x})=K'({\vec x'}),\qquad
A_i({\vec x})=A_i'({\vec x'}),
\ee
and the forms $B_i$ and $C_{ij}$ are defined by (\ref{DualFields}).

The functions $H^{-1}, K, A_i, B_i$ can be linearly superposed to
give different solutions of the D1-D5 system. These functions
depend only on $\vec x$; further $B_i$ is determined by $A_i$ as
described above.

\section{Unbound solution.}
\label{AppUnbound}
\renewcommand{\theequation}{D.\arabic{equation}}
\setcounter{equation}{0}

We wish to construct a solution of the D1-D5 system which
represents two  D1-D5 bound states, with opposite angular momenta
so that the total angular momentum is zero. We can either
construct the solution from two strings in the FP system and
perform dualities to the D1-D5 system, or work directly with the
chiral null models derived for the D1-D5 system in appendix B and
superpose two solutions. We will do the latter in this appendix.

Let us look at the simplest such superposition: we take a solution
with $Q_1,Q_5,a$ and add it to the solution with $Q_1,Q_5,-a$
(actually we will take an average to preserve the asymptotic value
of $H$). The resulting system is described by
\bea
H^{-1}=1+\frac{Q_5}{f_0},\qquad K=\frac{Q_1}{f_0},\qquad
A_i=0
\eea

We can rewrite the result of superposition in terms of the original spherical
coordinates $r,\theta,\phi,\psi$ (see (\ref{coord1}), (\ref{FlatCoord})):
\bea\label{D1D5App0}
ds^2&=&-\frac{{ f}_0}{\sqrt{f_1f_5}}(dt^2-d{ y}^2)+
\sqrt{f_1f_5}(\frac{dr^2}{r^2+{ a}^2}+d\theta^2)+
\sqrt{\frac{f_1}{f_5}}\sum_{i=1}^4 dz^i dz^i
\nonumber\\
&+&\sqrt{\frac{f_1}{f_5}}(1+\frac{Q_5}{f_{0}})
\left[r^2\cos^2\theta d\psi^2
+(r^2+{a}^2)
\sin^2\theta d\phi^2\right]\\
e^{2\Phi}&=&\frac{{ f}_5}{{ f}_1},\qquad
C^{(2)}_{ty}=-\frac{Q_1}{{ f}_1}, \qquad
C^{(2)}_{\phi\psi}=
\frac{Q_5(r^2+a^2)\cos^2\theta}
{{ f}_{0}}.
\eea

To compare this solution with metrics for infinite throat and for the
rotating D1--D5 branes we reduce the system (\ref{D1D5App0}) to six dimensions
($t,r,\theta,\phi,\psi,{ y}$) and look at the resulting metric in the
Einstein frame:
\bea\label{myToCompare}
ds_E^2&=&e^{-\frac{4\Phi}{6-2}}ds_6^2=-\frac{{ f}_0}{\sqrt{{ f}_1
{ f}_{5}}}(dt^2-d{ y}^2)
+\sqrt{{ f}_1 { f}_{5}}(\frac{dr^2}{r^2+a^2}+d\theta^2)
\nonumber\\
&+&\frac{\sqrt{{ f}_1 { f}_{5}}}{{ f}_0}
\left[r^2\cos^2\theta d\psi^2+
(r^2+a^2)\sin^2\theta d\phi^2\right]
\eea

We see that this metric has the form of a uniform throat only for
$r\gg a$; the geometry ends at $r\sim a$. Thus the resulting throat
has a length that reflects the value of $|a|$ of each component,
rather than the total angular momentum (which is zero). We will
examine the nature of the singularity at $r=0, \theta=\pi/2$ later
on in appendix \ref{AppThroat}.

\section{Throat geometry for  generic D1-D5 bound states}
\label{AppThroat}
\renewcommand{\theequation}{E.\arabic{equation}}
\setcounter{equation}{0}

We wish to understand the supergravity solution corresponding to a
general state out of the set of $\sim
e^{2\sqrt{2}\pi\sqrt{n_1n_5}}$  Ramond ground states of the D1-D5
system. While we have seen in Appendix \ref{AppGener} how to write  a metric
using the idea of chiral null models directly for the D1-D5 system,
we will proceed by first writing the metric for the FP system and
then taking its dual. The reason for this is that we  wish to make
solutions that correspond   to a single bound state of the D1-D5
system, rather than to a multicenter solution obtained from a
collection of D1-D5 bound states. In the FP language a single
bound state is given by the oscillations of a single string, and
we start with these solutions.

Consider a single fundamental string carrying a momentum wave with profile
$\vec G(v')$.  Then the solution (\ref{ChiralNull}) with
\bea\label{SingleString}
{H'}^{-1}({\vec x'},v')=1+\frac{Q'}{|{\vec
x'}-{\vec G}(v')|^2},
&&\qquad K'({\vec x}',v')=\frac{Q'|{\dot{\vec
G}}(v')|^2}{|{\vec x'}-{\vec G}(v')|^2}
\nonumber\\
A'_i({\vec x'},v')&=&-\frac{Q'{\dot{G}}_i(v')}{|{\vec
x'}-{\vec G}(v')|^2}
\eea
describes a fundamental string located
at ${\vec x'}={\vec G}(v')$.

As in \cite{lm3} we construct a superposition of such solutions
by smearing over the coordinate $y$ (or equivalently, by smearing
over $v$). This smearing arises because there are many
closely spaced strands of the string in the space $\vec x'$, and in
the classical limit
their smoothed out effect is just obtained by superposing the continuous
distribution given by the average over $v'$:
\bea
\langle H'^{-1}\rangle({\vec x'})&=&
\int_0^{L'}\frac{dv'}{L'}H^{-1}({\vec x'},v'),\qquad
\langle K'\rangle({\vec x'})=
\int_0^{L'}\frac{dv'}{L'}K({\vec x'},v'),\nonumber\\
\langle A'_i\rangle({\vec x'})&=&
\int_0^{L'}\frac{dv'}{L'}A'_i({\vec x'},v').
\eea
Note that we obtain the solution (\ref{FPnullCoor}) if we
perform this smearing on the profile  ${\vec G}(v')$:
\be
G_1(v')=a'\cos(\omega v'+\alpha),\qquad
G_2(v')=a'\sin(\omega v'+\alpha),\qquad G_3(v')=G_4(v')=0,
\ee

After duality transformations we get the D1-D5 solution
(\ref{D1D5Chiral}) with coefficient functions
\bea\label{Smearing}
\langle H^{-1}\rangle({\vec x})&=&1+\frac{Q}{L}\int_0^{L}
\frac{dv}{\sum (x_i-F_i(v))^2},\\
\langle K\rangle({\vec x})&=&\frac{Q}{L} \int_0^{L}\frac{\sum{\dot
F}_i{\dot F}_i dv}
{\sum (x_i-F_i(v))^2},\\
\langle A_i\rangle({\vec x})&=&-\frac{Q}{L}\int_0^{L} \frac{{\dot
F}_i dv}{\sum (x_j-F_j(v))^2}.
\label{SmearEnd}
\eea
(We will not need the explicit form of $\langle B_i\rangle$ and
$\langle C_{ij}\rangle$.)
>From these expressions one can see that the functions
$\langle H^{-1}\rangle$, $\langle K\rangle$, $\langle A_i\rangle$
are regular everywhere, except at points where ${\vec x}={\vec
F}(v)$ for some $0\le v<L$. Note that ${\vec F}$ is related to
${\vec G}$ by
\be
{\vec F}=\frac{g}{R\sqrt{V}}{\vec G}.
\label{rfour}
\ee

After dimensional reduction on the compact 4-manifold $M$ we get
the following 6-D Einstein metric for the D1-D5 system
\be\label{MGE1}
ds^2_E=\sqrt{\frac{H}{1+K}}\left[-(dt-A_idx^i)^2+(dy+B_idx^i)^2\right]
+ \sqrt{\frac{1+K}{H}}d{\vec x}\cdot d{\vec x} \ee with parameters
given by (\ref{Smearing})--(\ref{SmearEnd}).
By translational invariance we can set
\be
\int_0^L dv F_i(v)=0.
\ee
            We will also assume that singularity
in confined in the region with a typical size $a$, which means
that
\be F_i(v)F_i(v)\le a^2 \qquad\mbox{for all}\qquad v \ee
Then
for ${\vec x}^2\gg a^2$ we get:
\bea
\langle H^{-1}\rangle({\vec x})&\approx&1+\frac{Q}{{\vec x}^2},\\
\langle K\rangle({\vec x})&\approx&\frac{Q}{{\vec x}^2}
\frac{1}{L} \int_0^{L}|{\dot {\vec F}}|^2 dv, \eea \be \langle
A_i\rangle({\vec x})=O(({\vec x}^2)^{-3/2}),\qquad \langle
B_i\rangle({\vec x})=O(({\vec x}^2)^{-3/2})
\ee This leads to the leading approximation for the metric
(\ref{MGE1}):
\be\label{AsymSol1}
ds^2_{E}\approx
-\frac{1}{h}(dt^2-dy^2)+hd{\vec x}d{\vec x},
\ee
where
\bea\label{AsymSol2}
h&=&\left[\left(1+\frac{Q}{{\vec x}^2}\right)
\left(1+\frac{{\tilde Q}}{{\vec x}^2}\right)\right]^{1/2},\\
\label{DefQPrime}
{\tilde Q}&=&Q\frac{1}{L}\int_0^{L}|{\dot {\vec F}}|^2
dv.
\eea This solution can be trusted in the region ${\vec x}^2\gg
a^2$. We observe that in this approximation we get a geometry of
the D1--D5 system with $Q_5=Q$ and $Q_1={\tilde Q}$.

We now wish to define a parameter $\hat a$ which reduced to the
parameter $a$ for the special
geometries (\ref{D1D5}) but for a general throat measures the
effective size of the singularity (and thus
determines the effective length of the throat).
Thus we set
\be\label{DefineAHat}
{\hat a}\equiv\left[\frac{1}{L}\int_0^L |{\vec F}|^2 dv\right]^{1/2}
\equiv \left[\langle |{\vec F}|^2\rangle\right]^{1/2}.
\ee
Taking into account the relation between charges (\ref{DefQPrime}), we
can rewrite this expression as
\be\label{BraneSize} {\hat
a}=\sqrt{\frac{Q_1}{Q_5}}\left(\frac{\langle |{\vec F}|^2\rangle}
{\langle |{\dot{\vec F}}|^2\rangle}\right)^{1/2},
\ee

Similarly, in the dual FP system we have
\be\label{StringSize} {\hat
a}'=\sqrt{\frac{Q_P}{Q_w}}\left(\frac{\langle |{\vec G}|^2\rangle}
{\langle |{\dot{\vec G}}|^2\rangle}\right)^{1/2}, \ee
where $\vec F, \vec G$ are related as in (\ref{rfour})  and
\be\label{AasAPrime} {\hat a}'=\frac{{\hat a}R\sqrt{V}}{g},\qquad
\frac{Q_P}{Q_w}=\frac{Q_1}{Q_5}. \ee
In the FP system the string closes after $n_w=n_5$ windings around the circle
$y'$ of length $2\pi R'$, so we may write \be
G_i(v)=\sum_{n=1}^{\infty}C^{(n)}_i \cos\left(\frac{nv}{n_5
R'}+\alpha_{n,i}\right) \ee This leads us to the following
averages:
\bea
\langle |{\vec G}|^2\rangle&=&\frac{1}{2}\sum_{n=1}^{\infty}\sum_i
\left(C^{(n)}_i\right)^2,\\
\langle |{\dot{\vec G}}|^2\rangle&=&\frac{1}{2(n_5R')^2}
\sum_{n=1}^{\infty}\sum_i
n^2\left(C^{(n)}_i\right)^2.
\eea
If we define an ``average
harmonic'' ${\bar n}$ by
\be {\bar n}\equiv n_5R' \left(\frac{\langle
|{\dot{\vec G}}|^2\rangle}{\langle |{\vec G}|^2\rangle}\right)^{1/2},
\ee
then equation (\ref{StringSize}) becomes:
\be
{\hat a}'=\sqrt{\frac{Q_P}{Q_w}}\frac{n_5R'}{{\bar n}}.
\ee

Using the relations (\ref{AasAPrime}) and (\ref{DefCharge}) we find
for the D1-D5
system
\be
{\hat a}=\frac{\sqrt{n_1n_5}}{{\bar n}}\frac{g}{R\sqrt{V}}
\ee
We also define the generalization of the dimensionless parameter $\gamma$ (eqn.
(\ref{stwo}))  for the D1-D5 system:
\be\label{NewDefGaN}
{\hat\gamma}\equiv\frac{{\hat
a}R}{\sqrt{Q_1Q_5}}
\ee
which yields (using (\ref{DefCharge}))
\be\label{NewExpGaN}
{\hat\gamma}= \frac{1}{{\bar n}}
\ee

\section{Wave equation near the singularity.}
\renewcommand{\theequation}{F.\arabic{equation}}
\setcounter{equation}{0}
\label{AppWave}

Let us look at the massless Klein--Gordon equation in the  metric
(\ref{MGE1}); the coefficient functions in
the metric are given by (\ref{Smearing}) - (\ref{SmearEnd}).  We will assume
that the scalar field $\Phi$ does not
depend on
$y$ coordinate, and it depends on $t$ only through the factor
$\exp(-i\omega t)$:
\be
\Psi(x_i,t,y)=\exp(-i\omega t){\tilde\Psi}(x_i)
\ee
Then the wave equation becomes:
\be\label{waveEqn}
-\omega^2
\left[-\frac{1+K}{H}+A_iA_j\delta^{ij}\right]\Psi
-i\omega\d_iA_i\Psi-
2i\omega A_i\d_i\Psi+\d_i\d^i\Psi=0
\ee

>From the expressions (\ref{Smearing}) - (\ref{SmearEnd}) one can see that
the functions $\langle
H^{-1}\rangle(\vec x)$,
$\langle K\rangle(\vec x)$, $\langle A_i\rangle(\vec x)$ are regular
everywhere, except at points
where ${\vec x}={\vec F}(v)$ for some $0\le v<L$.

Consider the wave equation in the neighborhood of the point $\vec
x_0=\vec F(v_0)$ on the
singularity. We assume that
(i) the singularity curve has no
self-intersections, so that the direction of the
curve is well defined at $\vec x$ and
(ii) $|\dot{\vec F}(v_0)|\ne
0$.  Consider the 3-dimensional plane
normal to the singular curve at $\vec x_0$, and parameterize the
points in this plane by $\vec y$:
\be
x_i=F_i(v_0)+y_i,~~~\dot F_i(v_0)y_i=0.
\ee
and we will look only at the displacements orthogonal to the singularity:
\be
F_i(v_0)y_i=0.
\ee

Since $y$ is a compact direction, all functions $F_i(v)$ are periodic
($F_i(v+L)=F_i(v)$). We then get
\be
\d_i \langle A_i\rangle=0.
\ee
Consider the potential  term in (\ref{waveEqn})
\be
\omega^2
\left[\frac{1+K}{H}-A_iA_j\delta^{ij}\right]\Psi
\label{ythree}
\ee
Each of the coefficient functions $H^{-1}, K, A_i$ diverge as
$y^{-2}$ near the singular curve. It will be
important for us that the combination occurring in the potential has a
much softer divergence ($\sim
y^{-1}$) rather than the naively suggested $\sim y^{-4}$. We work
this out below, and give an
explanation of this softening in Appendix \ref{AppSelfInt}.

We have
\bea\label{TermsToCancel}
&&\left[1+\langle K\rangle\right]\langle H^{-1}\rangle-
\langle A_i\rangle\langle A_i\rangle=\nonumber\\
&&\qquad\left(\frac{Q}{L}\right)^2\int_0^{L}\int_0^{L}
\frac{dvdv'}{({\vec x}-{\vec F}(v))^2({\vec x}-{\vec F}(v'))^2}
\frac{({{\dot {\vec F}}(v)-{\dot {\vec F}}(v'))^2}}{2}\nonumber\\
&&\qquad+
\frac{Q}{L}
\int_0^{L}\frac{(1+{\dot F}_i{\dot F}_i) dv}
{({\vec x}-{\vec F}(v))^2}+1
\eea

We will consider only the curves without self intersection, then the main
contribution to this integral comes from the vicinity of the point $v_0$.
Let us evaluate the contribution from such vicinity to the first integral in
(\ref{TermsToCancel}):
\bea
I_1&\equiv&\left(\frac{Q}{L}\right)^2\int_0^{L}\int_0^{L}
\frac{dvdv'}{({\vec x}-{\vec F}(v))^2({\vec x}-{\vec F}(v'))^2}
\frac{({{\dot {\vec F}}(v)-{\dot {\vec F}}(v'))^2}}{2}\nonumber\\
&\approx&2\left(\frac{Q}{L}\right)^2
\int_0^{L}
\frac{dv({\dot{\vec F}}(v)-{\dot {\vec F}}(v_0))^2}
{{\vec y}^2+({{\vec F}}(v)-{ {\vec F}}(v_0))^2}
\int_0^{L}
\frac{dv'}
{{\vec y}^2+[{\dot{\vec F}}(v_0)]^2(v'-v_0)^2}\nonumber\\
&&\quad -2\left(\frac{Q}{L}\right)^2
\sum_i \int_0^{L}
\frac{dv({\dot{F}}_i(v)-{\dot F}_i(v_0))}
{{\vec y}^2+({{\vec F}}(v)-{ {\vec F}}(v_0))^2}
\int_0^{L}
\frac{dv'({\dot{F}}_i(v')-{\dot F}_i(v_0))}
{{\vec y}^2+({{\vec F}}(v)-{ {\vec F}}(v_0))^2}\nonumber\\
&\approx&2\left(\frac{Q}{L}\right)^2
\frac{\pi}{|{\dot{\vec F}}(v_0)|\sqrt{{\vec y}^2}}
\int_0^{L}
\frac{dv({\dot{\vec F}}(v)-{\dot {\vec F}}(v_0))^2}
{({{\vec F}}(v)-{ {\vec F}}(v_0))^2}
\nonumber\\
&&\quad -2\left(\frac{Q}{L}\right)^2
|{\ddot{\vec F}}(v_0)|^2 \left[\int_0^{L}
\frac{dv(v-v_0)}
{{\vec y}^2+({{\vec F}}(v)-{ {\vec F}}(v_0))^2}\right]^2
\eea

           The integral in the square brackets
behaves as $\log ({\vec y}^2)$, and thus the leading asymptotics for $I_1$ is
\be\label{NewF9}
I_1=\frac{C_1(v_0)}{|\vec y|}+o(|\vec y|^{-1}),\qquad
C_1(v_0)=2\left(\frac{Q}{L}\right)^2
\frac{\pi}{|{\dot{\vec F}}(v_0)|}
\int_0^{L}
\frac{dv({\dot{\vec F}}(v)-{\dot {\vec F}}(v_0))^2}
{({{\vec F}}(v)-{ {\vec F}}(v_0))^2}
\ee
In the same fashion we will get the asymptotics of the second integral in
(\ref{TermsToCancel}):
\bea\label{IRes}
I_2&\equiv&\frac{Q}{L}
\int_0^{L}\frac{(1+{\dot F}_i{\dot F}_i) dv}
{({\vec x}-{\vec F}(v))^2}=\frac{C_2(v_0)}{|\vec y|}+o(|\vec y|^{-1})\\
\label{NewF11}
C_2(v_0)&=&\frac{\pi Q}{L}(1+|{\dot{\vec{F}}}(v_0)|^2)
\frac{1}{|{\dot{\vec{ F}}(v_0)|}}
\eea

Thus we find that
the expression (\ref{TermsToCancel}) behaves near the singularity as
\be\label{AddPAs}
\left[1+\langle K\rangle\right]\langle H^{-1}\rangle-
\langle A_i\rangle\langle A_i\rangle\sim
\frac{C(v_0)}{|{\vec y}|}
\ee
(with $C(v_0)>0$.)

To analyze the equation (\ref{waveEqn}) we will also need the asymptotic
behavior of
$A_i$, which can be easily extracted from (\ref{SmearEnd}) and (\ref{IRes}):
\be\label{AddAAs}
\langle A_i\rangle\sim \frac{C_3(v_0){\dot F}_i(v_0)}{|{\vec y}|},\qquad
C_3(v_0)=-\frac{\pi Q}{L}\frac{1}{|{\dot{\vec{ F}}(v_0)|}}
\ee

In the four dimensional space $x_1,x_2,x_3,x_4$ we have a vector ${\vec y}$
which parameterizes a three dimensional subspace orthogonal to
${\dot{\vec F}}(v_0)$ and coordinate $z$ which measures the
distance along the curve. The flat metric can be rewritten in terms of these
four coordinates:
\be\label{NewF14}
d{\vec x}d{\vec x}=d{\vec y}d{\vec y}+|{\dot{\vec F}(z)}|^2dz^2=
dr^2+r^2[d\theta^2+\sin^2\theta d\phi^2]+|{\dot{\vec F}(z)}|^2dz^2.
\ee
Here we have introduced the spherical coordinates $r,\theta,\phi$ in the
three dimensional space of ${\vec y}$.

Then substituting (\ref{AddPAs}) and
(\ref{AddAAs}) into the equation (\ref{waveEqn}), we get in
the
leading order in $r=\sqrt{\vec{y}\cdot \vec{y}}$ and $z$:
\be\label{SingWaveEqn}
\frac{C(z)\omega^2}{r}\Psi-
\frac{2i\omega C_3(z)}{r}\d_z\Psi+\frac{1}{|{\dot{\vec{F}}}(z)|}
\d_z\left(\frac{1}{|{\dot{\vec{F}}}(z)|}\d_z\Psi\right)
+r^{-2}\d_r(r^2\d_r\Psi)+
\frac{1}{r^2}\Delta_{\theta,\phi}\Psi=0
\ee
The variables $\theta$ and $\phi$ in this equation separate and we will look
for the solution in the form:
\be
\Psi(t,r,z,\theta,\phi)=\exp(-i\omega t)Y_{lm}(\theta,\phi)\Phi(r,z),
\ee
Let us first consider the case $l>0$. In this case in the vicinity of $r=0$ we
get an approximate equation:
\be
r^{-2}\d_r(r^2\d_r \Phi)-\frac{l(l+1)}{r^2}\Phi=0.
\ee
This is a Schroedinger equation with repulsive potential and the particle is
reflected from the barrier.

In the case of the S wave ($l=m=0$) it is convenient to introduce a new
coordinate $\rho=r^{1/2}$. Then equation (\ref{SingWaveEqn}) becomes
\be
C(z)\omega^2 \Phi-2i\omega C_3(z)\d_z \Phi+
\rho^2\d_z\left(\frac{1}{|{\dot{\vec{F}}}(z)|}\d_z\Phi\right)
+\rho^{-3}\d_\rho(\rho^3\d_\rho \Phi)=0.
\ee
Assuming that $z$ derivatives of $\Phi$ are bounded:
\be
|\d_z \Phi|\le B_1|\Phi|, \qquad |\d^2_z \Phi|\le B_2|\Phi|,
\ee
we get the asymptotic equation near $\rho=0$:
\be
\rho^{-3}\d_\rho(\rho^3\d_\rho \Phi)+\Lambda\Phi=0
\ee
This is just the equation for the three dimensional spherical wave which
moves in the constant potential.
Such a wave  reflects from $\rho=0$ after a finite number of
oscillations and in a finite time.
In the next Appendix we estimate the time spent near the singularity
and show that it is less
than (or at most comparable) to the time spent reaching $r\sim a$
from the start of the throat.

\section{Null geodesics near the singularity.}
\renewcommand{\theequation}{G.\arabic{equation}}
\setcounter{equation}{0}
\label{AppGeodes}

In this appendix we will look at  null geodesics near the
singularity, and estimate the time which a particle traveling along such
geodesics spends in the vicinity of the singularity.

To analyze the geodesics we consider the Hamilton-Jacobi equation
\be\label{DefHJ}
g^{\mu\nu}\frac{\d S}{\d x^\mu}\frac{\d S}{\d x^\nu}=0
\ee
in the background (\ref{MGE1}). We will assume that the particle does not
move in the $y$ direction:
\be
\frac{\d S}{\d y}=0.
\ee
Then the Hamilton-Jacobi equation (\ref{DefHJ}) becomes:
\be\label{HJEqn}
\left[-\frac{1+K}{H}+A_iA_j\delta^{ij}\right]\left(\frac{\d S}{\d t}\right)^2
+2A_i\frac{\d S}{\d t}\frac{\d S}{\d x^i} +
\frac{\d S}{\d x^i}\frac{\d S}{\d x^i}=0.
\ee
As in the Appendix \ref{AppWave} we will go from coordinates $x^i$ to the
coordinates $z,r,\theta,\phi$ using (\ref{NewF14}). We will only study the
geodesics which approach the singularity in the radial direction. For such
geodesics the equation (\ref{HJEqn}) becomes:\footnote{Note that
$A_i$ diverges at
$r= 0$ so there may be a residual contribution from the middle term
in (\ref{HJEqn}) even though the geodesic is radial to leading order
as $r\rightarrow 0$.
We do not expect this to qualitatively change the analysis, and so
ignore such a potential
contribution to (\ref{HJEqn}).}
\be\label{EqnHJ1}
\left[-\frac{1+K}{H}+A_iA_j\delta^{ij}\right]\left(\frac{\d S}{\d t}\right)^2
+\left(\frac{\d S}{\d r}\right)^2=0.
\ee
Since coefficients in this equation do not depend on $t$, the solution has
the form:
\be\label{HJAct}
S(t,r)=\omega t+{\tilde S}(r).
\ee
Near the singularity we can use the approximation (\ref{AddPAs}), which allows
us to rewrite (\ref{EqnHJ1}) as
\be\label{EqnHJ2}
\frac{\d {\tilde S}}{\d r}\approx \omega\frac{\sqrt{C_1(z)+C_2(z)}}{\sqrt{r}},
\ee
where $C_1(z)$ and $C_2(z)$ are given by (\ref{NewF9}) and (\ref{NewF11}).
For the
generic singularity we have:
\be\label{NewC1}
C_1(z)\approx 2\left(\frac{Q}{L}\right)^2
\frac{\pi}{|{\dot{\vec F}}(z)|}~L
\frac{\langle|{\dot{\vec F}}|^2\rangle}{\langle|{{\vec F}}|^2\rangle}=
\frac{2\pi Q_5}{L}\frac{1}{|{\dot{\vec F}}(z)|}\frac{Q_1}{{\hat a}^2}
\ee
Here we have used (\ref{DefQPrime}), (\ref{DefineAHat}) and the definitions of
$Q_1$ and $Q_5$: $Q_5=Q$, $Q_1={\tilde Q}$. From eqn (\ref{DefQPrime}) we can
see that if $Q_1\sim Q_5$, then $|{\dot{\vec F}}(z)|\sim 1$, and comparing
(\ref{NewF11}) with (\ref{NewC1}), we observe that $C_1(z)\gg C_2(z)$ for
${\hat a}\ll Q_1^{1/2}$.

Equation (\ref{EqnHJ2}) becomes
\be
\frac{\d {\tilde S}}{\d r}\approx \frac{\omega}{\sqrt{r}}
\left(\frac{2\pi}{L}\frac{1}{|{\dot{\vec F}}(z)|}
\frac{Q_1Q_5}{{\hat a}^2}\right)^{1/2}
\ee
Integration of this equation gives the expression for the action (\ref{HJAct})
\be
S(t,r)=\omega t+2\omega\sqrt{r}
\left(\frac{2\pi}{L}\frac{1}{|{\dot{\vec F}}(z)|}
\frac{Q_1Q_5}{{\hat a}^2}\right)^{1/2}
\ee
To get the travel time from $r=r_0$ to $r=0$ we differentiate this action with
    respect to $\omega$:
\be\label{SingDelT}
\Delta t_{sing}=2\left(\frac{2\pi}{L}
\frac{r_0}{|{\dot{\vec F}}(z)|}
\frac{Q_1Q_5}{{\hat a}^2}\right)^{1/2}=
\frac{2}{\pi}\left(\frac{2\pi}{L}
\frac{r_0}{|{\dot{\vec F}}(z)|}\right)^{1/2}\Delta t_{SUGRA},
\ee
where
\be
\Delta t_{SUGRA}=\pi\frac{\sqrt{Q_1Q_5}}{{\hat a}}
\ee
is the travel time from the start of the throat to  $r= \hat a$ and back.

To determine the value of $L$ we take the ratio of (\ref{BraneSize}) and
(\ref{StringSize}):
\be
\frac{{\hat a}}{{\hat a}'}=\left(\frac{\langle |{\vec F}|^2\rangle}
{\langle |{\dot{\vec F}}|^2\rangle}\right)^{1/2}
\left(\frac{\langle |{\dot{\vec G}}|^2\rangle}
{\langle |{{\vec G}}|^2\rangle}\right)^{1/2}=\frac{L}{L'}.
\ee
At the last step we used the fact that the profiles ${\vec F}$ and
${\vec G}$ are related by a simple rescaling. Using the expression for
${\hat a}'$ from (\ref{AasAPrime}) and the effective length for the FP system:
$L'=2\pi n_5R'$, we get:
\be
L=\frac{2\pi R'gn_5}{R\sqrt{V}}=\frac{2\pi gn_5}{R}=\frac{2\pi Q_5}{R}
\ee
Substituting this value in (\ref{SingDelT}) and replacing
$|{\dot{\vec F}}(z)|$ by the average value
\be
\langle |{\dot{\vec F}}|^2\rangle^{1/2} =\sqrt{\frac{Q_1}{Q_5}},
\ee
we get
\be
\frac{\Delta t_{sing}}{\Delta t_{SUGRA}}=
{2\over \pi}\left(\frac{r_0 R}{(Q_1Q_5)^{1/2}}\right)^{1/2}
\ee
Taking $r_0\sim {\hat a}$ and using (\ref{NewDefGaN})and
(\ref{NewExpGaN}), we get:
\be
\frac{\Delta t_{sing}}{\Delta t_{SUGRA}}\sim {2\over \pi}\sqrt{\hat\gamma}=
\frac{2}{\pi\sqrt{\bar n}}
\ee
Since $\bar n \ge 1$, we see that the time spent near the singularity
is of the order or smaller
than the estimate $\Delta t_{SUGRA}$. Note that for the metrics
(\ref{MaldToCompare}) we have performed
the exact computation for the travel time $\Delta t_{SUGRA}$ (eqn.
(\ref{timeSUGRA}));  this time
includes all effects of approaching the singularity and returning back.

\section{Length of the singularity.}
\renewcommand{\theequation}{H.\arabic{equation}}
\setcounter{equation}{0}
\label{AppenLen}

Consider the FP system, and look at the length of the singular
curve in the space $\vec x'$. We assume that the coupling is
weak, so that the metric is flat; the location of the string will
give a localized singularity representing the physical location
of the string.  From  the energy of oscillations we get:
\be
\frac{n_P}{R'}=T\int |{\dot{\vec G}}|^2 dy'
\ee
Then the length of the
string is
\be
D'=\int
dy\sqrt{|{\dot{\vec G}}|^2}\le {L'}^{1/2}\left[\int |{\dot{\vec
G}}|^2 dy'\right]^{1/2}= \sqrt{\frac{n_P
L'}{R'T}}=2\pi\sqrt{N\alpha'}
\ee
where
$T=\frac{1}{2\pi\alpha'}$.
Note that equality is attained only if all the vibrations of the F
string are in
the same harmonic.

Let us map this to the D1-D5 system, and consider this system also at
weak coupling so that spacetime
is flat everywhere except the singularity. Then  we get for the
length of the singular
curve
\be D=D'\frac{g\alpha'^{3/2}}{R\sqrt{V}}\le 2\pi\sqrt{N\alpha'}
\frac{g\alpha'^{3/2}}{R\sqrt{V}}=2\sqrt{2}\frac{\sqrt{NG^{(6)}}}{R}
\ee
We can then write for the `area' occupied by the singularity in the
$\vec x, y$ space
\be
Area = 2\pi R D \le     4\pi\sqrt{2}\sqrt{NG^{(6)}},
\ee
where we again note that equality is attained only if all the
component strings in the
       microstate are of equal length (this is equivalent
       to all vibrations being in the same harmonic for the dual FP system).

\section{Singularity curves having self intersections}
\label{AppSelfInt}
\renewcommand{\theequation}{I.\arabic{equation}}
\setcounter{equation}{0}

Consider the potential term (\ref{ythree}) that we found in the
scalar wave equation near the singular curve of the
D1-D5 geometry. We had found that this term was much less singular
than would appear from the behavior of its
individual factors.  This softening of the singularity was essential
for the fact that the wave reflects back in a finite
time from the singularity. Let us look at the FP system, and
investigate the essential reason for this softening as well
as the situations where the wave equation may become more singular.

Suppose we have a single strand of a fundamental string oscillating
with some profile $\vec G(v')$.  Then we find,
using (\ref{SingleString}), that
\be
{\cal P}\equiv K'(H'^{-1}-1)-A'_iA'_j\delta^{ij} =0
\ee
If however we have two strands of the string (whether joined  into
one string or arising from two
separate strings) then
\be
{\cal P}=\frac{Q'_1Q'_2}{|{\vec x}'-{\vec G}_1(v')|^2
|{\vec x}'-{\vec G}_2(v')|^2}
|{\vec G}_1(v')-{\vec G}_2(v')|^2,
\label{yfive}
\ee
which is nonzero unless the profiles ${\vec G}_1(v')$ and ${\vec G}_2(v')$
are the same.

The quantity ${\cal P}$ differs from the potential (\ref{ythree})
only by terms that are
comparatively nonsingular.  Now consider the multiwound string. If
the singular curve it produces in the space
$\vec x'$ has no self-intersections, then neighboring strands along
this curve have almost the same profile $\vec
G(v')$. In that case the quantity (\ref{yfive}) is comparatively
regular,  and thus we see that the softening of the
potential in the wave equation can be traced back to the vanishing of
${\cal P}$ for a single strand.

But if the singular curve has self-intersections, then at the same
point in the space $\vec x'$ we have strands with
quite different profiles  $G_1(v'), G_2(v')$. The denominators in
(\ref{yfive}) now cause ${\cal P}$ to be large, and
the potential in the wave equation is correspondingly more singular.

Since the singular curve is a 1-D hypersurface in the 4-D space $\vec
x'$, the generic singular curve has no self
intersections. But since simple examples may actually have such
intersections, we examine two such cases in this
Appendix.

\bigskip

(i)\q The `unbound state' solution (\ref{D1D5App0}) was constructed by
superposing two bound states, each of which had a
singularity on the same circle of radius $a$.  Let us look at the 
dual FP system.
Since
the rotation was oppositely directed in the two components,
we find at each point along the singular curve a pair of strands with
different profiles $G_1(v'), G_2(v')$. Thus each
point of the singular curve gives a more singular potential in the
wave equation than the generic $\sim 1/|\tilde y|$
(eqn. (\ref{SingWaveEqn})).

We now find that geodesics that go radially into  any point  with
$r=0, \theta=\pi/2$ do not return back in a
finite time $t$. The divergence in this time of flight is however
only logarithmic: if we separate the two components
of the D1-D5 solution by  a distance $\delta r$ then the time of flight to
the singularity  again becomes finite and of order
$\sim \Delta t_{CFT}\log {\delta r\over a}$. But now note that each
bound state in the solution had some mass $M$,
and thus can be localized in the space $\vec x$ only to an accuracy
$|\delta \vec x|\sim 1/M$. If we set $Q_1\sim
Q_5$ for the moment, and work out the effect of this fluctuation,
then we find that
\be
\Delta t_{SUGRA}\sim \Delta t_{CFT} \log j
\ee
where $j$ is the angular momentum of each of the bound states in the
solution.  We will not explore the meaning
of this logarithm further, but just note that such logs have appeared
before in relating CFT and supergravity
quantities \cite{km}.

\bigskip

(ii)\q As a second example we write down an explicit solution for the
single bound state where the singularity is a
straight line: this will be the limiting case where the ellipse in
figure \ref{figEllStr} degenerates and the angular momentum
becomes zero.

Consider the string of the FP system oscillating in the direction $x_1$:
\be\label{F1F2Dabl}
G_1(v')=a'\cos(\omega v'+\alpha),\qquad G_2(v')=G_3(v')=G_4(v')=0.
\ee
Averaging over $\alpha$ in the usual manner we get the classical
solution (\ref{D1D5Chiral}) with coefficient functions
\bea\label{eqn50}
H^{-1}&=&\left(1+\frac{Q_5}{2r}\left[
\frac{2(r^2-z^2+a^2)+2\sqrt{(r^2-z^2+a^2)^2+4z^2r^2}}
{(r^2-z^2+a^2)^2+4z^2r^2}\right]^{1/2}\right)\\
\label{eqn51}
K&=&\frac{Q_1}{ra^2}\left(-2r+\left[2(r^2-z^2+a^2)+
2\sqrt{(r^2-z^2+a^2)^2+4z^2r^2}\right]^{1/2}\right)\\
A_i&=&0.
\eea

At each point of the singularity we have two differently moving
strands of the string arising from the
two sides of the degenerating ellipse. Correspondingly we find that
there is a logarithmic divergence in the time of
flight to each point on the singularity; in addition there is a
stronger singularity at the endpoints $x_1=\pm a$ since
here $|\dot {\vec G}(v')|$ vanishes as well, and so the density of
points  along the singular curve (\ref{PoinDens}) diverges .

\end{document}